\documentclass[12pt]{article}

\usepackage[utf8]{inputenc}
\usepackage[T1]{fontenc}
\usepackage{amsmath,amssymb}
\usepackage{physics}
\usepackage{hyperref}
\usepackage{graphicx}
\usepackage{xcolor}
\usepackage[margin=2.5cm]{geometry}
\usepackage[round,authoryear]{natbib}

\title{\textbf{What Are We Talking About\\When We Talk About Nuclear Reactions}}
\author{Gregory Potel \\ Departamento de F\'isica Aplicada III, Universidad de Sevilla, Sevilla, Spain \\ \texttt{gpotel@us.es}}
\date{}

\begin{document}
  
\maketitle
\tableofcontents 

\newpage

% ============================================================
\section{Introduction}
\label{sec:intro}
Let us start by addressing heads-on the question implicit in the title of this Chapter, formulating it in the simplest possible way: what is a nuclear reaction? The answer is deceptively simple: a nuclear reaction is a process in which two nuclei interact and exchange energy, momentum (linear and/or angular), and/or mass (in the form of nucleons or clusters of nucleons)\footnote{We will focus here on reactions involving two nuclear species. The extension to more than two bodies is conceptually straightforward but technically very challenging: a very explicit instance of the saying `easier said than done'. We will also assume that we are in the non-relativistic regime, and the dynamical aspects of our discussion will be driven by the Schrödinger equation.}. However this simple definition belies the rich variety and complexity of the field of nuclear reactions theory. Ultimately, this complexity owes to the nature of the nuclear spectrum, which determines the available ways in which the nucleus as a quantum many-body system can absorb and dissipate energy. Within this context, it should come as no surprise that the density of states plays an important role, similar to the heat capacity in condensed matter physics.   

The atomic nucleus is a unique self-bound finite quantum many-body system, and sheer scientific curiosity  alone (aside from the fact that nuclear reactions are at the basis of our understanding of very important astrophysical processes, as well as of a variety of societal applications) justifies the theoretical interest in nuclear reactions. But there is also a practical side to the story: nuclear reactions are the main tool for probing the structure of nuclei and validating theoretical models of the nucleus. The essence of the use of nuclear reactions as a probe of nuclear structure is to account for the quantitative correlation between the observed exchange of energy, momentum, and mass, and the underlying structure of the colliding nuclear species. Although the considerations to be presented in this Chapter are of rather general nature, we will highlight here the role of nuclear reactions as a probe and an experimental tool for addressing our understanding of the nucleus. 

Within this context, this introductory Chapter will try to briefly address
\begin{enumerate}
 \item the basic physical ingredients that determine the nuclear spectrum and the associated energy scales;
 \item  the connection between the observed experimental results, typically in the form of cross sections, and the underlying nuclear structure presented in the previous point.
 \end{enumerate}
 The nature and scope of the present Chapter is to provide a roadmap of the field of nuclear reactions theory, and to set the stage for the more detailed discussions in the companion Chapters. The aim is to provide a conceptual framework that will allow the reader to understand how the different reaction theory approaches fit together, and how they relate to the underlying physics of nuclear reactions. The reader will find in other chapters in this volume detailed discussions about the methods addressing two seemingly quite distinct phenomena: \emph{direct} and \emph{compound} nuclear reactions. We hope that this brief Chapter will help them realize that whenever we are talking about either of these, we are  talking about the same thing: \emph{nuclear reactions}. 

\section{General properties of the nucleus}
\label{sec:properties}
\subsection{Some basic facts}\label{S2.1}
The nucleus can be viewed as a self-bound
quantum liquid with a sharp edge, defining a tight  surface. The radius of this drop can be inferred experimentally from electron scattering experiments, from which it is found that it grows with mass number roughly
as
\begin{equation}
  R \approx r_0 A^{1/3}, \qquad r_0 \approx 1.2\ \text{fm}.
  \label{eq:radius}
\end{equation}
This  simple scaling law  implies nuclear-matter saturation: since
$A \propto R^3$, the nucleon density\footnote{A more accurate value of this so-called saturation density is $\rho_0 \approx 0.16$~fm$^{-3}$, obtained from the charge density of $^{208}$Pb measured in elastic electron scattering~\citep{deVries1987}.}
\begin{equation}
  \rho_0 = \frac{A}{\frac{4}{3}\pi R^3} = \frac{3}{4\pi r_0^3}
  \approx 0.14\ \text{fm}^{-3}
  \label{eq:density}
\end{equation}
is very stable for all nuclei, essentially independent of $A$. Exceptions to this rule, like exotic halo nuclei, are therefore quite spectacular and call for special dedicated study and explanations \citep{Tanihata}.

The average binding energy per
nucleon,
\begin{equation}
  B/A \approx 8\ \text{MeV},
  \label{eq:BA}
\end{equation}
can be experimentally inferred from mass measurements, and it is also rather constant for $A \gtrsim 20$ stable nuclei. This
near-constancy is itself a consequence of saturation, Eq.~\eqref{eq:density}:
each nucleon interacts with only its near neighbors (the number of which is essentially constant for all nuclei), so the total binding
energy grows linearly with $A$ rather than as $A^2$ (the rate of growth of the number of pairs of nuclei), as it would for a
long-range two-body force. Although this number suffers large variations as one moves away from the valley of stability, we will use it as a reference and representative value for our estimates.

\subsection{The nuclear spectrum}
With the basic experimental facts gathered in the previous section, let us now make  simple estimates of some  energy scales and    of the nature of the nuclear spectrum.

For this, we need a dynamical model, and we'll start with  the simplest one: the free Fermi gas.  In this model, nucleons are treated as non-interacting fermions confined in a potential well, and their energy levels are filled according to the Pauli exclusion principle.
Filling $A$ non-interacting nucleons into plane-wave states up to a Fermi momentum $k_F$ gives $\rho_0 = g k_F^3/6\pi^2$, so that\footnote{This connection between the Fermi momentum and the nuclear density is strictly valid only in the thermodynamic limit $A\to\infty$, $R\to\infty$ while $A/R^3$ remains finite. For a Fermi gas confined in a finite volume, the relation is modified and depends on the shape of the container (binding potential well). However, we shouldn't be concerned about this in the context of this low-resolution picture.}
\begin{equation}
  k_F = \left(\frac{3\pi^2\rho_0}{2}\right)^{1/3} \approx 1.3\ \text{fm}^{-1},
  \qquad
  \varepsilon_F = \frac{\hbar^2 k_F^2}{2m} \approx 35\ \text{MeV},
  \label{eq:epsF}
\end{equation}
using $\rho_0$ from Eq.~\eqref{eq:density}, and where the degeneracy $g = 4$ reflects the fact that each single-particle state is doubly degenerate in spin and isospin. 

We are now ready to estimate the depth of the effective potential well confining the nucleons within the nuclear volume: 
\begin{equation}
  V_0 = \varepsilon_F + B/A \approx 35 + 8 \approx 43\ \text{MeV}.
  \label{eq:V0}
\end{equation}
in good agreement — within the $\sim 10\%$ expected of so simple an
estimate — with the value $V_0 \approx 45$~MeV used later in this section
to obtain the shell-model oscillator spacing $\hbar\omega_0$
(Eq.~\eqref{eq:sp}), itself consistent with the empirical value
$41\,A^{-1/3}$~MeV. 

In order to advance in our understanding of the nuclear spectrum, it will be useful to consider two complementary pictures of the nucleus, each of which captures a different aspect of its structure and dynamics. 

\subsubsection{The nucleus as a Fermi gas.}
In the first picture, the nucleus is a gas of $A$ fermions interacting
through a complex, effectively random $A$-body interaction. The hallmark of
this limit is a high density of many-body states, driven by a combinatorial
explosion: the number of ways to distribute a fixed excitation energy
$E^*$ among $A$ nucleons as particle-hole excitations grows extremely
rapidly with $A$. The statistical properties of the resulting spectrum
are well described by Random Matrix Theory, specifically the Gaussian
Orthogonal Ensemble (GOE)~\citep{Weidenmuller2009}: the energy levels repel each other, the
eigenstates look like random superpositions of basis states, and the
fluctuations of the $S$~matrix from one energy to the next are those
of a chaotic quantum system. In this limit the nucleus has, loosely
speaking, a large heat capacity: any energy deposited by an incoming
projectile is rapidly shared among many degrees of freedom and the
system retains no memory of how it was formed. This is the realm of
\textit{compound reactions}. 

This combinatorial explosion can be made concrete as a counting problem
outright, assuming that, for the sake of our estimations, the complexity of the system of interacting nucleons can be reduced by postulating a common mean field. Picture $A$ fermions occupying
single-particle levels equally spaced by $\varepsilon_0$: in the ground
state the lowest $A$ levels are filled according to the Pauli principle, and an excitation simply promotes
some fermions to higher, originally empty levels, leaving holes behind. A
given excited configuration is specified by how far each promoted particle
has moved up and each hole sits below the Fermi level, and its total
excitation energy $E^*$ is just the sum of these individual jumps, each an
integer multiple of $\varepsilon_0$. Counting the number of such
configurations at fixed $E^*$ is therefore exactly the classical problem of
writing $E^*/\varepsilon_0$ as a sum of positive integers — a problem
already solved by Euler~\citeyearpar{Euler1753}, two centuries before its nuclear
application. The exact count grows combinatorially fast with $E^*$;
extracting its precise asymptotic growth rate for large $E^*$ is a more
technical exercise.

The one physical input this counting problem needs — the single-particle
spacing $\varepsilon_0$ — is already fixed by the two basic facts of
Sect.~\ref{sec:properties}. The mean-field well of depth $V_0$ and radius
$R_0=r_0A^{1/3}$ derived there confines its levels into major shells
spaced by $\hbar\omega_0$ (Eq.~\eqref{eq:sp} below); but each shell is far
from a single level — it is degenerate, holding $\sim N^2$ nearly
coincident orbitals, with $N$ the shell's principal quantum number. Since
the cumulative particle count up to shell $N_F$ scales as
$A\sim N_F^3$ (each successive shell adds $\propto N^2$ particles), the
shell index at the Fermi surface itself grows only as $N_F\propto A^{1/3}$,
and the degeneracy of that topmost shell scales as $N_F^2\propto A^{2/3}$.
The levels actually relevant to the counting argument are the individual
members of this degenerate set, spaced by
$\varepsilon_0\sim\hbar\omega_0/A^{2/3}\propto A^{-1}$ — the same scaling,
with an explicit normalization, as $\varepsilon_0=1/g(\varepsilon_F)$ with
the free-Fermi-gas density of states $g(\varepsilon_F)$ used below. For
$A=56$ this gives $\varepsilon_0\sim0.4$~MeV: individual levels barely half
an MeV apart, out of which the counting below builds up thousands of
many-body levels per MeV. A fully microscopic version of this same
counting problem — resolving the actual finite-well single-particle
spectrum level by level, together with its associated symmetry-energy
splitting, rather than an idealized equidistant ladder — was carried out
by hand for $A=20$ by Critchfield and Oleksa~\citeyearpar{CritchfieldOleksa1951};
for the order-of-magnitude estimates of this chapter the equidistant
approximation above is sufficient.

These are exactly the two ingredients the brute-force counting above
needs — but not in a form that stays practical. Even restricted to a
single light nucleus and $25$~MeV of excitation, Critchfield and Oleksa's
direct enumeration already demanded extensive hand computation; scaled up
to the medium and heavy nuclei and higher excitation energies of interest
in reaction theory, exact counting is simply hopeless.

What makes the problem tractable is trading exact enumeration for a
controlled approximation. The route pioneered by Bethe~\citeyearpar{Bethe1936}
and developed into a systematic technique by Bloch~\citeyearpar{Bloch1954} treats
the same counting problem thermodynamically: rather than enumerate
configurations one by one, one constructs the partition function of the
Fermi gas and extracts the level density from it by a saddle-point
(steepest-descent) evaluation — asymptotically exact in the limit of many
quanta. This machinery is developed in full in Ericson's classic
review~\citeyearpar{EricsonAdvPhys1960}; we do not reproduce it here, where we will just use the result.

That result is that the level density grows exponentially in the square root
of the excitation energy, at a rate set by the single-particle level
spacing near the Fermi surface. Its remaining shortcomings are
quantitative rather than qualitative — real nuclear levels are definitely not perfectly
equidistant nor free of pairing and shell structure — and are repaired by
two standard empirical corrections, systematically fit across the whole
mass table by Dilg \textit{et al.}~\citeyearpar{Dilg1973}: a back-shift of the
excitation energy, entering as $U-\Delta$ rather than $U$ itself and
absorbing the extra binding from pairing and shell closures, and a
spin-cutoff parameter $\sigma$ controlling how the total state count is spread across angular momenta, so that the density
of distinct levels (as opposed to individual $M$-substates) becomes
\begin{equation}
  \rho(U) = \frac{1}{12\sqrt{2}\,\sigma\,a^{1/4}}\,
  \frac{e^{2\sqrt{a(U-\Delta)}}}{(U-\Delta+t)^{5/4}},
  \qquad U-\Delta = at^2 - t,
  \label{eq:backshift}
\end{equation}
where $t$ is the (Lang--Le Couteur) nuclear temperature, fixed implicitly
by the second relation. We apply this below.

Three typical, non-magic cases from Dilg \textit{et al.} give the order of
magnitude that matters for what follows, quoted here in terms of $D_0$,
the mean spacing between consecutive $s$-wave neutron resonances at
$S_n$ — directly (although, in general, not easily!) measurable, and  related to $\rho(S_n)$ through
Eq.~\eqref{eq:backshift}: for the medium-light nucleus $^{58}$Fe
($S_n=10.0$~MeV), $a\approx6$~MeV$^{-1}$ and
$D_0\approx1.5$~keV; for medium-mass $^{120}$Sn ($S_n=9.1$~MeV),
$a\approx13$~MeV$^{-1}$ and $D_0\approx70$~eV; and for the heavy, fissile
$^{236}$U ($S_n=6.6$~MeV), $a\approx25$~MeV$^{-1}$ and
$D_0\approx0.5$~eV. Level spacings at $S_n$ shrink from the keV scale to
well below 1~eV across this range — by the time a compound nucleus forms
in a medium or heavy nucleus, its levels are not just numerous but
overlapping, which is exactly the regime Sect.~\ref{sec:compound}
models statistically.

\subsubsection{The nucleus as an object}\label{S2.2.2}
So far we have modeled the nucleus as a gas of independent particles
moving in a common mean field — but real nucleons interact with each
other, through a residual interaction left over once that mean field has
been subtracted out. The simplest possible hypothesis about this residual
interaction is that it is, for all practical purposes, random: that its
matrix elements in the shell-model basis behave as the entries of a random
matrix drawn from the Gaussian Orthogonal Ensemble (GOE), already invoked
qualitatively above. This is not an empty assumption — it makes sharp,
testable predictions, and around $S_n$ they hold up well: nearest-neighbor
level spacings follow the Wigner distribution (level repulsion, rather
than the Poissonian statistics of uncorrelated levels), and partial widths
follow the Porter-Thomas distribution, both hallmarks of GOE
statistics~\citep{Weidenmuller2009,Weidenmuller2010}. 

And yet we know the nucleon-nucleon interaction is not random. Quite the
opposite: it ``conspires'' in a specific way to produce a handful of striking, unmistakably
ordered, emergent properties. Nuclei are liquid drops, with a genuine
surface (Sect.~\ref{S2.1}). That drop
can be deformed, acquiring a definite shape and a definite orientation in
space. And, more subtly but no less concretely, nuclei can be
superfluid — deformed not in ordinary space but in \textit{gauge} space.
All three are instances of spontaneously broken symmetry: the nuclear
Hamiltonian is rotationally and gauge invariant, yet its mean-field
solutions need not be, and when they are not, a privileged shape or a
privileged phase emerges~\citep{Broglia2021}. It is precisely this symmetry
breaking that turns a system of otherwise uncorrelated fermions into a
genuine \textit{object}.

Each broken symmetry comes with its own emergent, collective degree of
freedom, conjugate to the corresponding constant of motion: the nuclear
radius for the surface; the ratio of the principal axes of inertia and
their orientation in space for the deformation; the pairing gap and the
gauge angle for superfluidity~\citep{Broglia2021}. A crucial point in the context of the present discussion is that each
of these degrees of freedom behaves as an \textit{elementary mode of
excitation} in its own right, and therefore carries its own spectrum,
embedded within the full nuclear spectrum. Two features distinguish this
collective spectrum sharply from the Fermi-gas spectrum built up in the
previous paragraphs. First, its quanta are bosonic — phonons of
vibration, rotation, pairing — not fermionic particle-hole pairs. An important exception is single-particle motion, a degree of freedom associated with the emergence of a collective mean field self-consistent with the nuclear density. Second,
and decisively, there are far fewer collective degrees of freedom than
there are nucleons: a handful of shape and gauge coordinates, against $A$
individual particles. This is the realm of \emph{direct} reactions. The resulting spectrum is correspondingly far
sparser than the combinatorially exploding Fermi-gas spectrum — let us now
explore the implications of this difference.

\textit{Single-particle energies.}
A nucleon bound in a spherical potential well of depth $V_0 \approx 45$~MeV
and radius $R_0 = r_0 A^{1/3}$ (with $r_0 \approx 1.2$~fm) experiences a
harmonic confining force with frequency:
\begin{equation}
  \hbar\omega_0 = \hbar c\sqrt{\frac{V_0}{m_N c^2 r_0^2}}\,A^{-1/3}
  \approx 36\,A^{-1/3}\ \text{MeV},
  \label{eq:sp}
\end{equation}
in good agreement with the empirical shell-model value $41\,A^{-1/3}$~MeV.
For $A = 56$ this gives $\hbar\omega_0 \approx 10$~MeV; for $A = 208$,
$\approx 7$~MeV. These are the energies of giant resonances, whose
$A^{-1/3}$ scaling~\citep{Bohr1975} follows directly from
Eq.~\eqref{eq:sp}. The \textit{spacing} of individual single-particle
levels near the Fermi surface is smaller by a factor $\sim A^{2/3}$
(the number of degenerate orbits per shell), placing it in the
1--5~MeV range for medium and heavy nuclei.

\textit{Vibrational energies.}
The surface term of the Weizsäcker formula~\citeyearpar{Weizsacker1935},
$E_s = a_s A^{2/3}$ with $a_s \approx 17.8$~MeV, defines a surface
tension $\sigma = a_s/(4\pi r_0^2) \approx 1$~MeV~fm$^{-2}$.\footnote{This
value need not be taken from mass systematics: by the same short-range,
coordination-number argument of Sect.~\ref{sec:properties} that
explains why $B\propto A$, a nucleon at the surface loses, by a
solid-angle argument, roughly half its near neighbors relative to the
bulk, at an areal density $n_s\sim\rho_0^{2/3}$ (Eq.~\eqref{eq:density}).
This gives $\sigma\approx\frac12\rho_0^{2/3}(B/A)\approx1.1$~MeV~fm$^{-2}$,
within $10\%$ of the value quoted above.}
For a quadrupole ($\lambda = 2$) oscillation of the nuclear surface, the
restoring-force stiffness is $C_2 = (\lambda-1)(\lambda+2)R_0^2\sigma
\propto A^{2/3}$ (proportional to the surface area, as it must be for a
surface-tension restoring force) and the kinetic-energy inertia for
irrotational flow is $B_2 = \frac{3m_N r_0^2}{8\pi}
A^{5/3}$~\citep{Bohr1975}. Combining
them gives the classical surface-vibration frequency (Rayleigh mode of
the liquid drop):
\begin{equation}
  \hbar\omega_2^{\rm vib} = \hbar\sqrt{\frac{C_2}{B_2}}
  \approx \frac{37}{\sqrt{A}}\ \text{MeV}.
  \label{eq:vib}
\end{equation}
For $A = 56$ this yields $\approx 4.9$~MeV; for $A = 150$,
$\approx 3.0$~MeV. The observed low-lying quadrupole phonons in
vibrational nuclei fall at $0.5$--$2$~MeV — a factor of $2$--$10$
lower than Eq.~\eqref{eq:vib}. The discrepancy is physical: nuclear
superfluidity increases the effective inertia well above the
irrotational-flow value (nucleon pairs do not contribute to the
irrotational current), reducing $\omega_2^{\rm vib}$ accordingly.
The high-energy collective quadrupole mode (the giant quadrupole
resonance), driven by the mean-field restoring force rather than
surface tension, follows the single-particle scale:
$E_{\rm GQR} \approx 63\,A^{-1/3}$~MeV~\citep{Bohr1975}.

\textit{Rotational energies.}
A deformed nucleus rotating as a rigid body of moment of inertia
$\mathcal{I}_{\rm rig} = \frac{2}{5}m_N r_0^2 A^{5/3}$ has a first
excited ($2^+$) state at:
\begin{equation}
  E_{2^+}^{\rm rot} = \frac{3\hbar^2}{\mathcal{I}_{\rm rig}}
  = \frac{15\hbar^2}{2m_N r_0^2 A^{5/3}}
  \approx \frac{216}{A^{5/3}}\ \text{MeV}.
  \label{eq:rot}
\end{equation}
For $A = 100$ this gives $\approx 100$~keV; for $A = 200$,
$\approx 32$~keV. In practice, nuclear superfluidity again reduces the
moment of inertia to $\sim 30$--$60\%$ of the rigid-body value, pushing
$E_{2^+}$ up by a factor of $\sim 2$. The empirically observed first
$2^+$ states in well-deformed heavy nuclei lie in the range
$50$--$200$~keV~\citep{Bohr1975} — precisely the regime predicted by
Eq.~\eqref{eq:rot} with this correction. Rotational energies are thus
the \textit{smallest} energy scale in the nuclear spectrum.

\textit{Pairing-vibrational energies.} Here the natural energy scale is
set experimentally, by the odd-even mass staggering: nuclei with an odd
$A$ are systematically less bound than their even-$A$ neighbors, by an
amount identified with the pairing gap~\citep{BohrMottelson1969},
\begin{equation}
  \Delta_{\rm pair} \approx \frac{12}{\sqrt{A}}.
  \label{eq:gap}
\end{equation}
The same gap shows up directly in the spectrum, as the energy separating
the ground state of an even-even nucleus from the lowest states of its
odd-$A$ neighbors, and it has a natural interpretation as a correlation
energy: the extra binding gained by forming a Cooper pair, over and above
what the same two nucleons would contribute uncorrelated. In principle a
restoring force and an inertia in gauge space, analogous to $C_2$ and
$B_2$ above, could be defined and combined into a vibration frequency the
same way; in practice this is considerably more involved than the shape
case, and we do not attempt it here. As a concrete, precisely-determined
example instead: the harmonic pairing-vibrational energy in the
$^{208}$Pb region, fit directly to the mass systematics of neighboring
Pb isotopes, is $\hbar\omega \approx 2.494$~MeV~\citep{Broglia2021}.

The contrast with the Fermi-gas estimate of the previous paragraph is
stark. The collective spectrum in the same nucleus ($^{56}$Fe,
$E^* \sim 0$--$10$~MeV) consists of a handful of discrete, identifiable
states: ground-state band members, specific phonon excitations,
giant-resonance peaks, single-nucleon orbitals. A window of 1~MeV anywhere in this range
contains at most a few levels, compared to the thousands predicted by
the Fermi-gas model. It is this \textit{coexistence} of the two
pictures in the same nucleus — one sparse and collective, the other
dense and chaotic — that gives rise to the rich variety of nuclear
reaction mechanisms we discuss in the remainder of this chapter.

\subsubsection{The synthesis}\label{S2.2.3}
Of course, real nuclei live between these two extremes. Both pictures
are simultaneously operative: the same nucleus that supports a
well-defined ground-state rotational band also has, at higher excitation
energies, a dense quasi-continuum of many-body states with GOE
statistics. The character of a reaction — direct or compound — depends
on which face of the nucleus the projectile happens to engage, and
at what energy.

This qualitative picture can be made concrete with a single Hamiltonian.
Write $H = H_F + H_B + V_C$, where $H_F$ is the (effectively random,
GOE-distributed) fermionic Hamiltonian of the Fermi-gas picture above,
$H_B$ is the bosonic Hamiltonian of the collective degrees of freedom
just described — surface, shape, and pairing phonons, with single-particle
motion itself the one genuinely fermionic exception among these
``elementary modes'' — and $V_C$ couples the two. Such a partitioning is,
formally, no restriction at all: for \textit{any} choice of $H_F$ and
$H_B$ one can simply define $V_C \equiv H - H_F - H_B$ and the
decomposition is exact but empty. It becomes physically useful only when
$V_C$ is small enough that it does not mix different eigenstates of $H_B$
carrying the same quantum numbers — precisely the condition under which
the elementary modes above retain their identity as approximately good
excitations, rather than dissolving into the fermionic background. This
is the strategy underlying Nuclear Field Theory~\citep{Bes1983}.

\begin{figure}[htbp]
  \centering
  \includegraphics[width=0.9\textwidth]{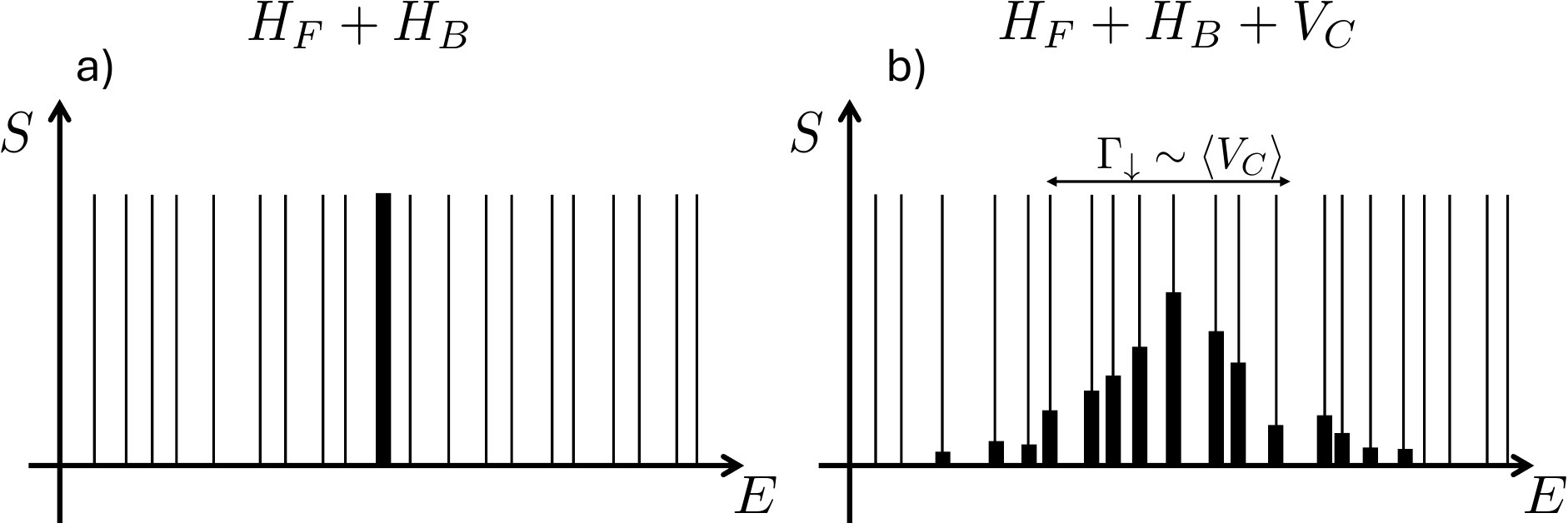}
  \caption{Schematic strength function $S(E)$ of a collective mode.
  (a) The spectrum of $H_F+H_B$: a single isolated eigenstate of $H_B$
  (thick line) embedded in the dense, near-continuous spectrum of $H_F$
  (thin lines). (b) The spectrum of $H_F+H_B+V_C$: the coupling fragments
  the collective strength over a band of $H_F$ eigenstates of width
  $\sim\langle V_C\rangle$, without destroying it. Over an energy range of the order of the spreading width $\Gamma_\downarrow$, the nuclear states retain a finite content (overlap) of the collective state, represented here by the length of the thick lines.}
  \label{fig:fragmentation}
\end{figure}

Figure~\ref{fig:fragmentation} makes the point pictorially. In (a), the
spectrum of $H_F + H_B$ alone: a dense, GOE-distributed fermionic
spectrum (thin lines) with a single isolated collective state of $H_B$
(thick line) superposed on it, carrying all of some observable's
strength $S$ at one sharp energy. In (b), once $V_C$ is switched on, that 
strength is not destroyed but redistributed: the collective state
fragments over a band of nearby fermionic eigenstates, spread across an
energy range of order $\langle V_C\rangle$ defining a spreading width $\Gamma_\downarrow$, while remaining recognizable
as a single physical entity smeared over many exact eigenstates rather
than lost among them~\citep{BertschBrogliaBortignon}. An early example was provided by Lane, Thomas, and Wigner, describing  what they called ``giant resonances''  associated with the mixing of single-particle states with  more complicated (compound) nuclear states~\citep{LaneThomasWigner}. In this particular case, $H_B$ is just the nucleon-nucleus kinetic energy plus the single-particle average potential.  

The aim of this chapter is to make this picture precise and to trace its
consequences for how we model nuclear reactions. We will show that the
distinction between direct and compound reactions has a clean mathematical
formulation in terms of the $S$~matrix, which can be decomposed into a
smooth energy-averaged part (the direct amplitude) and a rapidly
fluctuating part (the compound amplitude). From this decomposition the
modeling frameworks for each class of reactions emerge naturally. The
specific techniques — Hauser-Feshbach theory, the distorted-wave Born
approximation, coupled-channels methods, and their modern extensions —
are treated in depth in the companion chapters of this volume; here we
aim to exhibit the common conceptual skeleton that underlies all of them.
 
% ============================================================
\section{Scattering theory in a nutshell}
\label{sec:Smatrix} 

\subsection{The asymptotic scattering wavefunction}
\label{sec:GF}

We summarize here succinctly some standard results of quantum scattering
theory, associated with the scattering of two
nuclei $x$ and $A$; for a fuller treatment we refer the reader to 
standard
textbooks, e.g.~\citep{ThompsonNunes2009,Jackson1970,Messiah1995,Satchler1990}.
Throughout, $x$ is treated as a structureless particle, while the
internal structure of both $A$ and of the compound system $B\equiv A+x$
is kept in full. With this understanding, the Hamiltonian of the system
is\footnote{All operators below are understood to be non-local unless
stated otherwise; when it is useful to display the arguments of an
operator explicitly we write them just once, e.g.\
$\hat{A}(\mathbf{r})\equiv\hat{A}(\mathbf{r},\mathbf{r}')$, with no loss
of generality, since a local operator is simply the special case
$\hat{A}(\mathbf{r},\mathbf{r}')=A(\mathbf{r})\delta(\mathbf{r}-\mathbf{r}')$.}
\begin{equation}
  H = h_B(\mathbf{r},\xi) = T_x + h_A(\xi) + V(\mathbf{r},\xi),
  \label{eq:Hstruct}
\end{equation}
where $\xi$ collectively denotes the internal (spatial and spin)
coordinates needed to describe the structure of $A$, $\mathbf{r}$ is the
$x$-$A$ relative coordinate, $T_x$ the kinetic energy of relative motion,
$h_A$ the intrinsic Hamiltonian of $A$, and $V$ the interaction between $x$ and $A$.
The Schrödinger equation $(E-H)\Psi(\mathbf{r},\xi)=0$ rearranges into
\begin{equation}
  \bigl(E-T_x-h_A(\xi)\bigr)\,\Psi(\mathbf{r},\xi)
  = V(\mathbf{r},\xi)\,\Psi(\mathbf{r},\xi), 
\end{equation}
which suggests building the exact solution out of the incident wave
$\Psi_0$, satisfying the free equation $(E-T_x-h_A)\Psi_0=0$, and the
inverse of the unperturbed Hamiltonian,
\begin{equation}
  G_0(E) = \lim_{\eta\to0^+}\bigl(E-T_x-h_A+i\eta\bigr)^{-1},
\end{equation}
the unperturbed many-body Green's function, or propagator. The solution $\Psi_0$ to the homogeneous equation expresses the boundary condition selected by the experiment, in terms of the beam energy and incident direction.  The
infinitesimal $\eta$ is taken to zero only \textit{after} the inversion:
without it $G_0$ would be singular whenever $E$ coincides with a real
eigenvalue of $T_x+h_A$, leaving the inversion ill-defined; taking $\eta$
positive and vanishingly small further guarantees that the resulting
scattered wave is a genuinely outgoing spherical wave, enforcing the
correct asymptotic boundary
condition~\citep{Dickhoff2005,Messiah1995}. One verifies directly, by
applying $(E-T_x-h_A)$ to both sides, that the wavefunction defined by
the \textit{Lippmann-Schwinger equation}~\citep{Lippmann1950}
\begin{equation}
  \Psi = \Psi_0 + G_0(E)\,V\,\Psi
  \label{eq:LS}
\end{equation}
indeed solves the Schrödinger equation. This can equivalently be written
\begin{equation}
  \Psi = \Psi_0 + G(E)\,V\,\Psi_0,
  \label{eq:LSalt}
\end{equation}
with
\begin{equation}
  G(E) = \lim_{\eta\to0^+}\bigl(E-T_x-h_A-V+i\eta\bigr)^{-1}
\end{equation}
the total, fully-interacting many-body Green's function. The equivalence
of Eqs.~\eqref{eq:LS} and~\eqref{eq:LSalt} is  a standard result of scattering theory,
following from Dyson's equation relating $G_0$ and $G$,
\begin{equation}
  G = G_0 + G_0\,V\,G = G_0 + G\,V\,G_0,
  \label{eq:Dyson}
\end{equation}
so the choice between them is purely a matter of
bookkeeping~\citep{Dickhoff2005}: Eq.~\eqref{eq:LS} builds up the exact
wavefunction order by order in $V$ acting on the free Green's function,
while Eq.~\eqref{eq:LSalt} packages the entire perturbation series into a
single action of the exact resolvent $G$ on the free wave. In either
form, the first term on the right is the incident channel fixed by our
boundary condition, and the second is the scattered wave. The
unperturbed wavefunction satisfying the standard scattering boundary
conditions is
\begin{equation}
  \Psi_0(\mathbf{r},\xi) = \Phi_0(\xi)\,F(\mathbf{k}_x,\mathbf{r}),
  \label{eq:Psi0}
\end{equation}
with $F$ a free incident plane wave of momentum $\mathbf{k}_x$ and
$\Phi_0$ the ground state of $A$.\footnote{When both $x$ and $A$ are
charged, it is convenient to remove the Coulomb interaction from $V$ and
absorb it instead into the unperturbed Green's function,
$G_0=\bigl(E-T_x-h_A-Z_AZ_xe^2/r\bigr)^{-1}$, in which case $F$ becomes a
Coulomb function rather than a plane wave, without otherwise affecting
anything that follows.}

\textit{Elastic scattering.} We now expand the total wavefunction in the
complete set of eigenstates $\{\Phi_i(\xi)\}$ of $h_A$,
\begin{equation}
  \Psi(\mathbf{r},\xi;E) = \sum_i \Phi_i(\xi)\,\psi_i(\mathbf{r}),
  \qquad
  \bigl(\epsilon_i-h_A(\xi)\bigr)\Phi_i(\xi) = 0,
  \label{eq:expand}
\end{equation}
and project Eq.~\eqref{eq:LS} onto $\Phi_0$. This gives a one-body
Lippmann-Schwinger equation for the elastic ($i=0$) channel wavefunction,
\begin{equation}
  \psi_0(\mathbf{r}) = F(\mathbf{r}) + \langle\Phi_0|G(E)V|\Phi_0\rangle\,F(\mathbf{r}).
  \label{eq:elastic}
\end{equation}

\textit{Inelastic scattering.} Equation~\eqref{eq:elastic} generalizes
directly to the case where $A$ has been excited to a state $i$ of its
spectrum,
\begin{equation}
  \psi_i(\mathbf{r}) = \langle\Phi_i|G(E)V|\Phi_0\rangle\,F(\mathbf{r}).
  \label{eq:inelastic}
\end{equation}
The free wave $F$ does not appear on the right hand side of the equation above: as part of our boundary
condition, only the ground state of $A$ is present in the incident beam,
and the population of every inelastic ($i\neq0$) channel is due entirely
to the scattering process driven by $V$.

\textit{Cross section and the \textit{T} matrix.} The cross section
follows from the asymptotic ($\mathbf{r}\to\infty$) form of $\psi_i$,
which describes the system far from the interaction region, where the
detectors sit. Using the asymptotic form of the Green's function,
\begin{equation}
  \langle\Phi_i|G(\mathbf{r}\to\infty)V|\Phi_0\rangle
  = \mathcal{O}(k_i,\mathbf{r})\,\langle\psi_i\Phi_i|V|\Phi_0F\rangle,
\end{equation}
with $\mathcal{O}(k_i,\mathbf{r}) = -\dfrac{\mu}{2\pi\hbar^2}\dfrac{e^{ik_ir}}{r}$
the standard asymptotic outgoing wave of momentum
$k_i=\sqrt{2\mu(E-\epsilon_i)}/\hbar$ ($\mu$ the $x$-$A$ reduced mass),
Eqs.~\eqref{eq:elastic} and~\eqref{eq:inelastic} give
\begin{equation}
  \psi_i(\mathbf{r}\to\infty) = F(\mathbf{r}\to\infty)\,\delta_{i0}
  + \mathcal{O}(k_i,\mathbf{r})\,T_{i0},
  \qquad
  T_{i0} \equiv \langle\psi_i\Phi_i|V|\Phi_0F\rangle,
  \label{eq:fGV}
\end{equation}
which defines $T_{i0}$, an element of the \textit{T}-matrix connecting
entrance channel $0$ to exit channel $i$. Note that, unlike a
first-order perturbative amplitude, Eq.~\eqref{eq:fGV} is exact: $T_{i0}$
is built from the exact scattered wave $\psi_i$, not from any truncation
of a power series. The differential cross section is expressed in terms
of the T-matrix as~\citep{ThompsonNunes2009,Jackson1970}
\begin{equation}
  \frac{d\sigma_{i0}}{d\Omega} = \frac{\mu^2 k_i}{4\pi^2\hbar^4 k_0}\,|T_{i0}|^2.
  \label{eq:dsigma}
\end{equation}

% It is useful to generalize the T-matrix by making explicit the energies
% at which the initial ($\Phi_0F$) and final ($\psi_i\Phi_i$) states are
% evaluated — in the present context, the center-of-mass energy of the
% system — and by allowing $V$ itself to be energy-dependent, as it is,
% for instance, whenever it is identified with a complex, energy-dependent
% optical potential (Sect.~\ref{sec:optical}). With this generalization,
% the T-matrix entering Eq.~\eqref{eq:dsigma} is the \textit{full-on-shell}
% T-matrix,
% \begin{equation}
%   T_{i0}(E,E,E) = \langle\psi_i\Phi_i(E)|V(E)|\Phi_0F(E)\rangle,
% \end{equation}
% and it is also useful to define the \textit{half-on-shell} T-matrix,
% \begin{equation}
%   T_{i0}(E,E,E') = \langle\psi_i\Phi_i(E)|V(E)|\Phi_0F(E')\rangle,
%   \qquad E\neq E',
% \end{equation}
% and, more generally, the \textit{off-shell} T-matrix
% $T_{i0}(E,E',E'')=\langle\psi_i\Phi_i(E)|V(E')|\Phi_0F(E'')\rangle$ with
% $E,E',E''$ all distinct (not needed in what follows). Whenever a single
% energy argument is given, or none at all, the full-on-shell T-matrix is
% understood: $T_{i0}\equiv T_{i0}(E)\equiv T_{i0}(E,E,E)$.

% \subsection{From the \textit{T} matrix to the \textit{S} matrix}
% \label{sec:Sdef}

Let us now introduce the S-matrix, defined in terms of the partial wave expansion  of the T-matrix by
\begin{equation}
  S_l = 1 - \frac{i\mu k}{\pi\hbar^2}\,T_l,
  \label{eq:Sdef}
\end{equation}
in terms of which the elastic and total reaction
cross sections take the familiar compact form
\begin{equation} 
  \sigma_{\rm el} = \frac{\pi}{k^2}\sum_l(2l+1)|1-S_l|^2,
  \qquad
  \sigma_{\rm reac} = \frac{\pi}{k^2}\sum_l(2l+1)\bigl(1-|S_l|^2\bigr).
  \label{eq:xsec}
\end{equation}

Equation~\eqref{eq:xsec} generalizes easily to many open channels.
Labeling the incoming and outgoing (observed) channels by  $\alpha,\beta$ — each specifying the mass partition, the projectile and target internal states, and the quantum numbers of relative
motion — and resolving every channel by total angular momentum $J$ and
parity $\pi$, the cross section from $\beta$ to $\alpha$ becomes
\begin{equation}
  \sigma_{\alpha\beta} = \frac{\pi}{k_\beta^2}
  \sum_{J\pi}(2J+1)\,\bigl|S_{\alpha\beta}^{J\pi} - \delta_{\alpha\beta}\bigr|^2.
  \label{eq:multichannelT}
\end{equation}

\subsection{Born series and the distorted-wave Born approximation}
\label{sec:BornDW}

Equation~\eqref{eq:LS} can be iterated to give the wavefunction as a
perturbative expansion in powers of $V$, the \textit{Born series},
\begin{equation}
  \Psi = \Psi_0 + G_0(E)\,V\bigl(\Psi_0 + G_0(E)\,V(\Psi_0+\cdots)\bigr).
  \label{eq:Born}
\end{equation}
Its first-order term, $\Psi\approx\Psi_0+G_0(E)V\Psi_0$, is the
(plane-wave) first order Born approximation. In practice, however, this is
rarely the most convenient way to organize the expansion. One typically
introduces an auxiliary potential $U_0(\mathbf{r})$ (usually chosen to be a simple, solvable complex potential) and rewrites the
Schrödinger equation as
\begin{equation}
  \bigl(E-T_x-h_A(\xi)-U_0(\mathbf{r})\bigr)\Psi(\mathbf{r},\xi)
  = \bigl(V(\mathbf{r},\xi)-U_0(\mathbf{r})\bigr)\Psi(\mathbf{r},\xi),
\end{equation}
with solution expressible, exactly as before, in the two equivalent forms
\begin{equation}
  \Psi = \widetilde\Psi_0 + \widetilde G_0(E)\,(V-U_0)\,\Psi
  \;=\; \widetilde\Psi_0 + G(E)\,(V-U_0)\,\widetilde\Psi_0,
  \label{eq:DWform}
\end{equation}
where $\widetilde G_0=\lim_{\eta\to0^+}(E-T_x-h_A-U_0+i\eta)^{-1}$ and
$\widetilde\Psi_0(\mathbf{r},\xi)=\Phi_0(\xi)\widetilde F(\mathbf{r})$,
with the \textit{distorted wave} $\widetilde F$ the solution of
$(E-T_x-h_A-U_0)\widetilde F=0$. The first-order term of the
corresponding power series,
\begin{equation}
  \Psi \approx \widetilde\Psi_0 + \widetilde G_0(E)\,(V-U_0)\,\widetilde\Psi_0,
  \label{eq:DWBAfirst}
\end{equation}
is the first-order \textit{Distorted Wave Born Approximation} (DWBA),
widely used in practice~\citep{ThompsonNunes2009,Jackson1970}. The freedom to choose $U_0$ can be
used to make the matrix elements of $V-U_0$ as small as possible,
accelerating the convergence of the  power series.

\section{Nuclear reactions as a measurement apparatus}
\label{sec:probe}

In the previous Section, we have managed to express the asymptotic form of the wavefunction (and therefore the cross section observed at the
detector, which is the result of our experiment) as a functional of
the structure of the whole projectile-plus-target system, encoded
entirely in the Green's function and the interaction through the
M{\o}ller operator $\Omega=1+GV$ (Eq.~\eqref{eq:LSalt}).

 However, we are not quite done yet. The next step is
to make more explicit the connection between the cross section and the underlying nuclear structure of the composite
projectile-plus-target system. A detector is a
measuring apparatus: through its interaction with the system, its
pointer acquires a definite value — a cross section, associated with a
specific final channel. The system being measured is the composite
projectile-plus-target nucleus, which exists \textit{as a nucleus} only
in the small spatial region
where projectile and target overlap. This picture — an apparatus (the
open reaction channels, playing the role of the pointer) coupled to a
measured system (the compound nucleus, confined to the interaction
volume) — motivates partitioning the full Hilbert space
into two orthogonal subspaces: $P$ for the open channels (measuring device), $Q$
for the compound nucleus (measured system). We will implement this strategy following the formalism first developed by Feshbach~\citeyearpar{Feshbach1958,Feshbach1962}.

\subsection{The model space: the $P$ and $Q$ operators}
\label{sec:Feshbach}

Let us consider for simplicity a nucleon scattering off a target nucleus of $A-1$
nucleons,\footnote{We follow here the convention of these notes, in
which the compound system has $A$ nucleons; this differs by one unit
from the $(A+1)$-body notation used earlier in this chapter for the same
physical system.} forming the compound system of $A$ nucleons. The
$A$-body Schrödinger equation reads
\begin{equation}
  (H-E)|\Psi\rangle = 0.
\end{equation}
Let us single out $\Lambda$ reaction channels $c$, each corresponding to one
nucleon in the continuum and the remaining $A-1$ nucleons in a bound
state of the target, and define the projector onto this portion of the
Hilbert space~\citep{Feshbach1958,Feshbach1962},
\begin{equation}
  P = \sum_{c=0}^{\Lambda}\int_{E_c}^{\infty} dE'\;
  |\xi_c(E')\rangle\langle\xi_c(E')|,
  \label{eq:P}
\end{equation}
where $E_c$ is the threshold energy of channel $c$, and the channel
wavefunctions
\begin{equation}
  |\xi_c(E')\rangle = \chi_c(r;E')\otimes\Phi_c^{A-1},
  \qquad (E'-H_{PP})|\xi_c(E')\rangle=0,
  \label{eq:xic}
\end{equation}
describe the relative motion $\chi_c(r;E')$ of the nucleon with respect
to the target in its bound state $\Phi_c^{A-1}$. Similarly, define the
projector onto the bound states of the full $A$-body system,
\begin{equation}
  Q = \sum_{\mu=1}^{M} |\Phi^A_\mu\rangle\langle\Phi^A_\mu|,
  \qquad (QHQ-\lambda_\mu)|\Phi^A_\mu\rangle=0,
  \label{eq:Q}
\end{equation}
where $\{|\Phi_\mu^A\rangle\}$ are the (purely discrete) bound states of
the $A$-body system; $\Lambda$ and $M$ are assumed large enough for the
numerical accuracy of interest. We identify the projection $QHQ$ of the
Hamiltonian onto the $Q$-space with the intrinsic Hamiltonian of the
$A$-body nucleus, and further assume $P+Q=1$, $PQ=QP=0$, so that the
full Hamiltonian decomposes as
\begin{equation}
  H = H_{PP}+H_{PQ}+H_{QP}+H_{QQ},
\end{equation}
with $H_{PP}\equiv PHP$, $H_{PQ}\equiv PHQ$, $H_{QP}\equiv QHP$,
$H_{QQ}\equiv QHQ$. Projecting the Schrödinger equation onto $P$ and $Q$
gives the coupled equations
\begin{align}
  (E-H_{PP})|\Psi_P\rangle &= H_{PQ}|\Psi_Q\rangle,
  \label{eq:P-eq}\\
  (E-H_{QQ})|\Psi_Q\rangle &= H_{QP}|\Psi_P\rangle,
  \label{eq:Q-eq}
\end{align}
with $|\Psi_P\rangle\equiv P|\Psi\rangle$, $|\Psi_Q\rangle\equiv
Q|\Psi\rangle$.

We will not address here the technical question of the actual implementation of this partitioning of the Hilbert space. 
It is important to note that $H_{QQ}$ is a model Hamiltonian which encodes the structure of the $A$-body system assuming that it is a closed system, i.e., uncoupled to the open channels described by $H_{PP}$. A possible (but by no means unique) choice for $H_{QQ}$ is the shell-model Hamiltonian of the $A$-body system~\citep{MahauxWeidenmuller1969}.  However, in order to achieve a clean separation between the measuring device and the measured system, it will be useful to ensure that the different possible outcomes of the experiment (i.e., channel cross sections) do not interact directly with each other. In other words, we will assume that $H_{PP}$ is diagonal in channel space, and that the transition from an initial (elastic) channel 0 and a final channel $a$, both belonging to subspace $P$, is entirely due to the interaction with the subspace $Q$, as described by the $T$-matrix (\ref{eq:Texact}).

A possible way to implement this is to assume that $P$ spans the region of the Hilbert space associated with a projectile-target distance larger than the range of the strong interaction between the colliding species, while $Q$ spans the complementary (internal) region. In this case, $P$ and $Q$ spaces are only directly connected through the continuity requirements of the wave function, and $H_{QP}$ is the Bloch operator enforcing the continuity of the first derivative of the projectile-target relative wavefunction~\citep{Bloch}. With this choice, the expression (\ref{eq:T-Feshbach}) of the $T$-matrix is equivalent to the one derived from  $R$-matrix theory~\citep{LaneThomas,Descouvemont}.

\subsection{The optical potential}
\label{sec:optical}

From Eq.~\eqref{eq:Q-eq}, we can solve formally for $|\Psi_Q\rangle$,
\begin{equation}
  |\Psi_Q\rangle = G_Q H_{QP}|\Psi_P\rangle,
  \qquad
  G_Q = \frac{1}{E^+-H_{QQ}}, \quad E^+\equiv\lim_{\eta\to0^+}(E+i\eta),
\end{equation}
and substitute back into Eq.~\eqref{eq:P-eq} to obtain the effective
Schrödinger equation for the wavefunction in the $P$-space,
\begin{equation}
  \bigl(E-H_{PP}-\Sigma(E)\bigr)|\Psi_P\rangle = 0,
  \qquad
  \Sigma(E) \equiv H_{PQ}G_QH_{QP},
  \label{eq:P-eff}
\end{equation}
where $\Sigma(E)$ is the contribution to the effective Hamiltonian in $P$-space generated by coupling to the
$Q$-space.  A
single-particle optical potential 
\begin{equation}
V(r,r';E)=U_0(r,r')+U_{\rm pol}(r,r';E)
  \label{eq:OP}
\end{equation}
 for the elastic
channel (sometimes called the Generalized Optical Model
Potential~\citep{MahauxWeidenmuller1969}) is obtained by projecting
$\Sigma(E)$ onto the ground state of the $(A-1)$-nucleus target. It has
a static, real, energy-independent part,\footnote{Although often
assumed local, this potential is in general non-local; a conspicuous
source of non-locality is the antisymmetrization between the incoming
nucleon and the nucleons of the $(A-1)$-nucleus (the Fock term of the
mean field).}
\begin{equation}
  \langle\Phi_{A-1}^0|H_{PP}|\Phi_{A-1}^0\rangle = T + U_0(r,r'),
\end{equation}
and a dynamic, complex, energy-dependent part — $\Sigma(E)$'s own
elastic-channel matrix element,
\begin{equation}
  U_{\rm pol}(r,r';E) = \langle\Phi_{A-1}^0|\Sigma(E)|\Phi_{A-1}^0\rangle
  = \langle\Phi_{A-1}^0|H_{PQ}G_QH_{QP}|\Phi_{A-1}^0\rangle.
\end{equation}
This dynamic part — often called the polarization potential — can be
rewritten by expanding the Green's function in the eigenstates of
$H_{QQ}$ (the shell-model states),
\begin{equation}\label{eq:U-pol}
  U_{\rm pol} = \sum_\mu \frac{\langle\Phi_{A-1}^0|H_{PQ}|\Phi_\mu^A\rangle
  \langle\Phi_\mu^A|H_{QP}|\Phi_{A-1}^0\rangle}{E^+-\lambda_\mu}
  = \lim_{\eta\to0^+}\sum_\mu\frac{V_{0\mu}(r)V_{\mu0}(r')}{E^+-\lambda_\mu},
\end{equation}
where the coupling potentials are
$V_{0\mu}(r)\equiv\langle\Phi_{A-1}^0|H_{PQ}|\Phi_\mu^A\rangle=V_{\mu0}^*(r)$.

 The real, static part $U_0$ is the mean field in which
the projectile moves — closely related to the shell-model potential well
of Eq.~\eqref{eq:sp}, but now felt by a scattering rather than a bound
particle. The dynamic part $U_{\rm pol}$ is complex and energy-dependent:
its imaginary part is absorptive, removing flux from the elastic channel.

The energy-averaged version $\langle V(E)\rangle$ of the optical potential plays a paramount role in direct reactions (see Sect.~\ref{sec:direct}). When the energy averaging interval $I$ is large compared to the mean level spacing $D$ of the compound nucleus, the resulting  potential has a smooth energy dependence, and, in particular, it doesn't exhibit the rapid oscillations driven by the energy denominator of Eq.~\eqref{eq:U-pol}.

In practice, $\langle V(E)\rangle$ is often fit to elastic scattering data and parametrized in
Woods-Saxon form, $f(r)=[1+\exp((r-R)/a)]^{-1}$. Global parametrizations fitted across wide ranges of $A$ and
$E$, such as that of Koning and Delaroche~\citeyearpar{KoningDelaroche2003},
encode decades of elastic-scattering systematics and provide the
transmission coefficients $T_c$ (Eq.~\eqref{eq:Tc} below) that feed
directly into the Hauser-Feshbach formula, Eq.~\eqref{eq:HF} below — the
optical model is thus the common input shared by the compound and direct
aspects of this Chapter. 

\subsection{The \textit{T} matrix}
\label{sec:Tshell}

The T-matrix in the $P$-space is associated with the asymptotic
($r\to\infty$) behavior of $|\Psi_P\rangle$. Eliminating $|\Psi_Q\rangle$
between Eqs.~\eqref{eq:P-eq} and~\eqref{eq:Q-eq} gives the
Lippmann-Schwinger equation for $|\Psi_P\rangle$,
\begin{equation}
  |\Psi_P\rangle = |\xi_0\rangle + G_PH_{PQ}G_QH_{QP}|\Psi_P\rangle,
  \label{eq:PspaceLS}
\end{equation}
where $|\xi_0\rangle$ is the elastic channel wavefunction of
Eq.~\eqref{eq:xic}, satisfying $(E-H_{PP})|\xi_0\rangle=0$, and
$G_P=(E^+-H_{PP})^{-1}$ is the Green's function in the $P$-space with
outgoing-wave boundary conditions. 

The equation above tells, in mathematical language, an intuitively satisfying story: the state of the measuring device $|\Psi_P\rangle$ is a superposition of the unperturbed state $|\xi_0\rangle$ and the perturbation induced by its interaction with the measured system, $H_{PQ}G_QH_{QP}|\Psi_P\rangle$. This interaction has a (formally) simple four-step structure: the measuring device couples to the measured system through $H_{QP}$, the measured system propagates in its own space through $G_Q$, and then it couples back to the measuring device through $H_{PQ}$. The final crucial step is the propagation of the perturbed measuring device itself through $G_P$, which ensures that the outgoing wave has the correct asymptotic form, and that both pieces of the asymptotic wavefunction (the unperturbed and the scattered one) belong to the Hilbert space $P$ in which the results of the measuring apparatus are recorded. 
The T-matrix is then the factor
multiplying the outgoing spherical wave in the asymptotic form of the
scattered-wave (second term of Eq.~\eqref{eq:PspaceLS}, see
Sect.~\ref{sec:GF})).

 The practical utility of this form is, however, limited by the fact that it is an implicit equation for $|\Psi_P\rangle$, which appears on both sides. The next two subsections
show how to obtain both an approximate and an exact expression for the
T-matrix in terms of the unperturbed, initial channel wavefunction $|\xi_0\rangle$ prepared by the experimental conditions. The final goal is to express the T-matrix (and, therefore, the observed cross section, see Eqs. (\ref{eq:dsigma},\ref{eq:xsec})), as a matrix element between the initial and final states of the measurement device, i.e., of states in the $P$ partition of the Hilbert space.

\subsubsection{First order approximation to the T-matrix.} Approximating the exact wavefunction by
setting $|\Psi_P\rangle\to|\xi_0\rangle$ in the scattered-wave term of
Eq.~\eqref{eq:PspaceLS} gives the first order approximation,
\begin{equation}
  |\Psi_P\rangle \approx |\xi_0\rangle + G_PH_{PQ}G_QH_{QP}|\xi_0\rangle,
\end{equation}
from which the first order T-matrix follows directly,
\begin{equation}
  T_{0a}^{(0)} = \langle\xi_a|H_{PQ}G_QH_{QP}|\xi_0\rangle.
  \label{eq:TDWBA}
\end{equation}
The poles of this T-matrix — its resonances — are simply the eigenvalues
of $H_{QQ}$, i.e., the intrinsic (in the sense of decoupled from the channel space $P$) states of the compound nucleus. Aside from its simplicity, the interesting feature of this approximation is to make the connection between the observed cross sections and the underlying states of the system under study very explicit: the reaction associated with the observed channel $a$ will exhibit the features (e.g., peaks of the cross section at resonant energies) associated with the structure of the composite projectile + target system defined in the space $Q$. 

Let us also point out that there is an important distinction between this first order approximation and the DWBA  discussed in Sect.~\ref{sec:dwba}. While the T-matrix of the latter is defined as a matrix element  between the initial and final distorted waves, defined as solutions of an auxiliary (usually complex) potential, the T-matrix of Eq.~\eqref{eq:TDWBA} is defined as a matrix element between scattering states in the $P$-space solutions of the \emph{real} Hamiltonian $H_{PP}$. 

\subsubsection{Parametrization of the exact T-matrix in terms of energies and widths.} An exact expression follows by looking for an
effective propagator $\mathcal G_Q=(E^+-\mathcal H_{QQ})^{-1}$ such that
\begin{equation}
  |\Psi_P\rangle=|\xi_0\rangle+G_PH_{PQ}\mathcal G_QH_{QP}|\xi_0\rangle
  \label{eq:PspaceLSexact}
\end{equation}
reproduces Eq.~\eqref{eq:PspaceLS} exactly. To determine the effective
Hamiltonian $\mathcal H_{QQ}$, equate Eqs.~\eqref{eq:PspaceLS}
and~\eqref{eq:PspaceLSexact},
\begin{equation}
  G_PH_{PQ}\mathcal G_QH_{QP}|\xi_0\rangle
  = G_PH_{PQ}G_QH_{QP}|\Psi_P\rangle
  \;\Longrightarrow\;
  \mathcal G_QH_{QP}|\xi_0\rangle = G_QH_{QP}|\Psi_P\rangle,
\end{equation}
and use Eq.~\eqref{eq:PspaceLS} once more to eliminate $|\xi_0\rangle$ in
favor of $|\Psi_P\rangle$,
\begin{align}
  \mathcal G_QH_{QP}\bigl(|\Psi_P\rangle-G_PH_{PQ}G_QH_{QP}|\Psi_P\rangle\bigr)
  &= G_QH_{QP}|\Psi_P\rangle, \notag\\
  \mathcal G_Q\bigl(1-H_{QP}G_PH_{PQ}G_Q\bigr) &= G_Q, \notag\\
  \bigl(1-H_{QP}G_PH_{PQ}G_Q\bigr)\bigl(E^+-H_{QQ}\bigr) &= E^+-\mathcal H_{QQ}, \notag\\
  E^+-H_{QQ}-H_{QP}G_PH_{PQ} &= E^+-\mathcal H_{QQ},
\end{align}
where the second line follows because $H_{QP}|\Psi_P\rangle$ multiplies
both sides and the resulting operator identity must hold for the
arbitrary state $|\Psi_P\rangle$, and the third line uses
$G_Q=(E^+-H_{QQ})^{-1}$. We finally obtain the effective Hamiltonian in
the $Q$-space,
\begin{equation}
  \mathcal H_{QQ} = H_{QQ}+H_{QP}G_PH_{PQ},
  \label{eq:HQQbar}
\end{equation}
so that the exact T-matrix is
\begin{equation}
  T_{0a} = \langle\xi_a|H_{PQ}\mathcal G_QH_{QP}|\xi_0\rangle
  = \langle\xi_a|H_{PQ}\bigl(E-H_{QQ}-H_{QP}G_PH_{PQ}\bigr)^{-1}H_{QP}|\xi_0\rangle.
  \label{eq:Texact}
\end{equation}
Comparing Eqs.~\eqref{eq:TDWBA} and~\eqref{eq:Texact}, the exact
T-matrix has exactly the first order structure, but with the bare $H_{QQ}$
replaced by the dressed $\mathcal H_{QQ}$ of Eq.~\eqref{eq:HQQbar}: the
compound-nucleus states are shifted and broadened by their coupling,
through $H_{QP}G_PH_{PQ}$, out to the open channels and back.

Let us now make more explicit the connection of the T-matrix with the composite nucleus by expanding it in the eigenstates of $H_{QQ}$.
Inserting the $Q$-space completeness relation of Eq.~\eqref{eq:Q} into
Eq.~\eqref{eq:Texact} gives
\begin{equation}
  T_{0a}(E) = \sum_{\mu\nu} W_{a\mu}(E)\,
  \bigl[D(E)^{-1}\bigr]_{\mu\nu}\,W_{0\nu}(E), 
  \label{eq:T-Feshbach}
\end{equation}
where the coupling matrix elements
\begin{align}\label{eq:widths}
  \nonumber W_{a\mu}(E) &= \langle\xi_a(E)|H_{PQ}|\Phi_\mu^A\rangle, \\ 
  W_{0\nu}(E) &= \langle\Phi_\nu^A|H_{QP}|\xi_0(E)\rangle
\end{align}
 are directly
connected to the channel partial widths of the corresponding shell-model
states, and the level matrix $D(E)$ encodes the $Q$-space physics
dressed by the coupling to the continuum,
\begin{equation}
  D_{\mu\nu}(E) = (E-\lambda_\mu)\delta_{\mu\nu}
  - \langle\Phi_\mu^A|H_{QP}G_PH_{PQ}|\Phi_\nu^A\rangle.
\end{equation}
Expanding $G_P$ once more in the $P$-space channel states of
Eq.~\eqref{eq:P} gives
\begin{equation}
  \langle\Phi_\mu^A|H_{QP}G_PH_{PQ}|\Phi_\nu^A\rangle
  = \sum_{c=0}^{\Lambda}\int_{E_c}^\infty dE'\;
  \frac{W_{\mu c}(E')W_{c\nu}(E')}{E^+-E'},
\end{equation}
which, on separating real and imaginary parts with the
Sokhotski-Plemelj theorem, gives
\begin{equation}
  D_{\mu\nu}(E) = (E-\lambda_\mu)\delta_{\mu\nu}
  - \Delta_{\mu\nu}(E) + \frac{i}{2}\Gamma_{\mu\nu}(E),
  \label{eq:Dmatrix}
\end{equation}
with the real energy-shift matrix
\begin{align}
  \Delta_{\mu\nu}(E) &= \mathrm{P}\!\int dE'\,\sum_c\frac{
    W_{\mu c}(E')W_{c\nu}(E')}{E-E'},
\end{align}
where the principal-value integral above is responsible for
renormalizing the bare shell-model energies $\lambda_\mu$ — and the
width matrix
\begin{equation}
  \Gamma_{\mu\nu}(E) = 2\pi\sum_{c=0}^{\Lambda}W_{\mu c}(E)\,W_{c\nu}(E).
  \label{eq:Gamma}
\end{equation}
The exact poles of the T-matrix are complex: their real part is shifted
from the shell-model eigenvalues $\lambda_\mu$ by the eigenvalues of
$\Delta_{\mu\nu}$, and their width is given by the eigenvalues of
$\Gamma_{\mu\nu}$.

The effective $Q$-space Hamiltonian $\mathcal H_{QQ}$ of Eq.~\eqref{eq:HQQbar} is non-Hermitian, but, since it is symmetric because of time-reversal invariance, it has a complete set of right and left eigenvectors (biorthogonal basis)~\citep{Berggren1968} with complex eigenvalues
\begin{equation}
\mathcal E_\mu = \lambda_\mu + \Delta_\mu - i\Gamma_\mu/2=E_\mu - i\Gamma_\mu/2,
\end{equation}
 that are the complex poles of the T-matrix. The real part of $\mathcal E_\mu$ is the energy of the compound-nucleus eigenstate, shifted from the bare shell-model energy $\lambda_\mu$ by the real part of the self-energy $\Delta_\mu$, while the imaginary part is the width of the state, given by the imaginary part of the self-energy $\Gamma_\mu/2$.  We can then obtain a convenient parametrization of the T-matrix expanding over the biorthogonal basis,
\begin{equation}
  T_{0a}(E) = \sum_{\mu=1}^{M}\frac{g_{a\mu}(E)\,g_{0\mu}(E)}{E-E_\mu(E)+\frac{i}{2}\Gamma_\mu(E)},
  \label{eq:T-diagonal}
\end{equation}
where the $g_{a\mu}(E)$ are the widths expressed in the new basis. 
\section{Direct and compound nucleus reactions}
\label{sec:divide}

The  T-matrix of Eqs.~\eqref{eq:T-Feshbach} and \eqref{eq:T-diagonal} is a sum over the complex
poles of $\mathcal H_{QQ}$ — the dense, GOE-distributed compound-nucleus
spectrum of Sect.~\ref{sec:properties}, spaced by the mean level
spacing $D$ associated with the inverse of the level density. Nowhere in this expression is there yet any trace of the
coarse, collective physics of $H_B$ anticipated in
Sect.~\ref{sec:properties} and pictured in Figure~\ref{fig:fragmentation}:
the exact T-matrix was built by treating every state of $\mathcal H_{QQ}$
on the same footing, fermionic and bosonic alike. Within this context, we will try in this Section to  take over the task expressed at the end of Sect.~\ref{sec:properties} and show how the
direct and compound contributions can be disentangled.

\begin{figure}[htbp]
  \centering
  \includegraphics[width=0.75\textwidth]{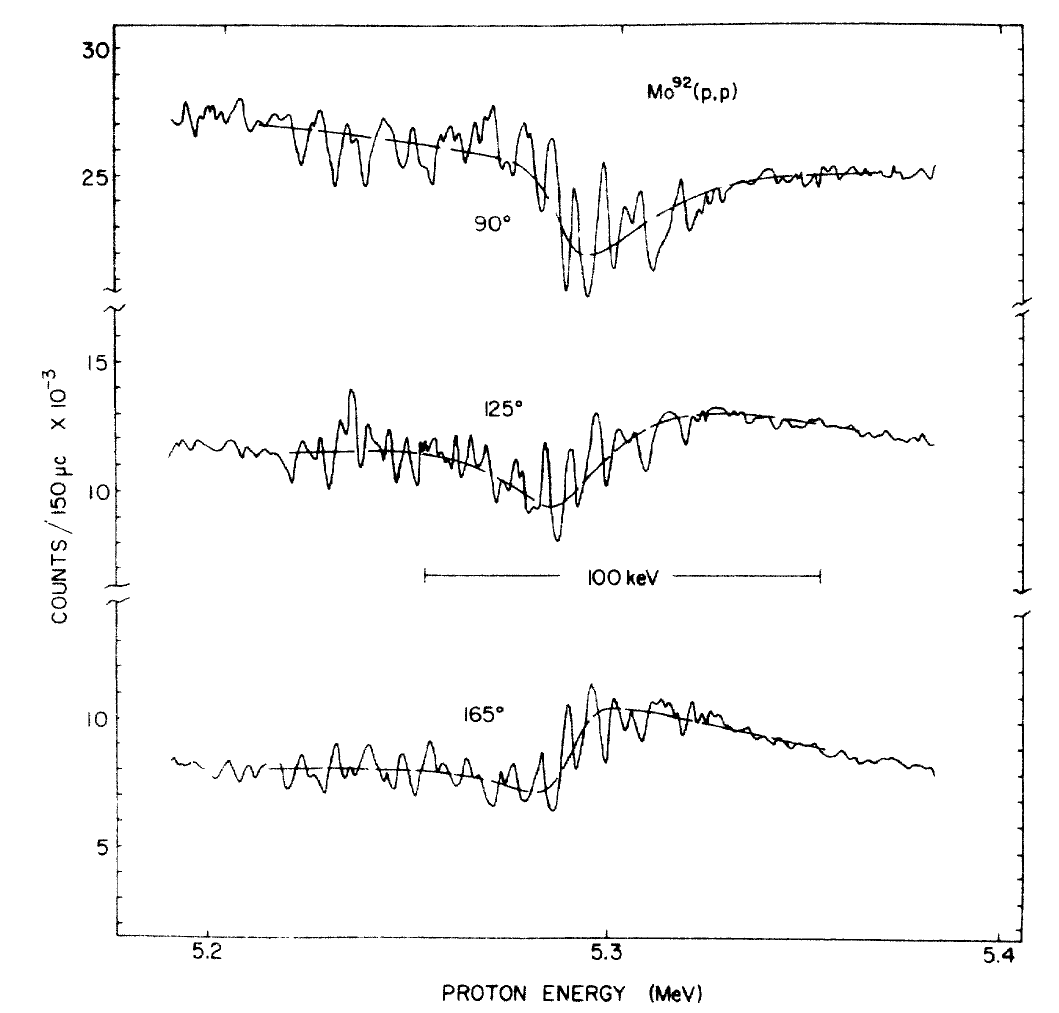}
  \caption{Proton elastic-scattering yield from $^{92}$Mo$+p$ at
  $\theta_{\rm lab}=90^\circ$, $125^\circ$, $165^\circ$, in the
  neighborhood of the $s$-wave isobaric analog resonance at
  $E_p=5.3$~MeV. Dense fine structure of width $\sim3$~keV is resolved
  at all three angles; the smooth curve is the single-level fit,
  $\Gamma=27$~keV, that only becomes visible once the fine structure is
  averaged over. Reproduced from Richard, Moore, Robson, and
  Fox~\citeyearpar{Richard1964}.}
  \label{fig:iar}
\end{figure}

In order to get a heuristic hint on how to proceed, let us look at Figure~\ref{fig:iar}. It shows the
$^{92}$Mo(p,p)$^{92}$Mo elastic-scattering excitation function at
$90^\circ$, $125^\circ$, and $165^\circ$ across the $s$-wave isobaric
analog resonance at $E_p=5.3$~MeV~\citep{Richard1964}. The isobaric
analog resonance is a collective state naturally described by $H_B$, i.e., a coherent excitation
built by isospin symmetry directly out of the target ground state. The raw data show only a dense forest of
$\sim3$~keV compound resonances; the broad, smooth resonance of width
$\Gamma=27$~keV (the drawn curve) emerges only once that fine structure
is averaged over. Energy-averaging has acted as a low-pass filter,
letting the coarse $H_B$ structure of Figure~\ref{fig:fragmentation}
survive while washing out the fine, $D$-spaced $H_F$ comb underneath it. 

\subsection{Energy averaging and the direct/compound decomposition}
\label{sec:decomp}

In practice, resonances in the T-matrix are so closely spaced that they
cannot be resolved individually with the available experimental
resolution. It is then useful to introduce the energy-averaged
T-matrix,
\begin{equation}
  \langle T_{0a}(E)\rangle = \int_{E-\Delta E/2}^{E+\Delta E/2}\,\rho(E-E')
  \,T_{0a}(E')\,dE',
\end{equation}
where $\rho(E-E')$ is a suitable, normalized distribution centered at
$E$. It is customary to choose a Lorentzian distribution for the
averaging process,
\begin{equation}
  \langle T_{0a}(E)\rangle = \frac{1}{\pi}\int_{-\infty}^{\infty}
  \frac{\Delta E/2}{(E-E')^2+(\Delta E/2)^2}\,T_{0a}(E')\,dE'
  = T_{0a}(E+i\Delta E/2).
\end{equation}
The last equality follows from the Cauchy integral formula, assuming that $T_{0a}(E)$ is analytic in the upper half-plane, and it tends to a constant as $E\to\infty$. The equivalence  of energy averaging a quantity over an interval $I$ and evaluating it at a complex energy $E+iI/2$ is a very useful property that will be used in the following.

 As we have argued, this averaging process is also useful in smoothing out the strength, highlighting the coarse structure of the collective states of $H_B$ while washing out the fine structure of the dense compound-nucleus spectrum of $H_F$. 
The energy-averaged S-matrix follows through the standard 
relation 
\begin{equation}
  S_{0a}(E) = \delta_{0a}-2\pi i\,T_{0a}(E),
  \qquad
  \langle S_{0a}(E)\rangle = \delta_{0a}-2\pi i\,\langle T_{0a}(E)\rangle.
  \label{eq:Snorm}
\end{equation}

Let us consider the elastic channel, with diagonal
S-matrix element $S_{00}(E)\equiv S$. The reaction and elastic cross
sections are, as in Eq.~\eqref{eq:xsec},
\begin{equation}
  \sigma_r = \frac{\pi}{k^2}\bigl(1-|S|^2\bigr),
  \qquad
  \sigma_{\rm el} = \frac{\pi}{k^2}|1-S|^2,
\end{equation}
and the experimentally observed cross sections are the energy-averaged
quantities $\langle\sigma_r\rangle$, $\langle\sigma_{\rm el}\rangle$,
obtained by replacing $|S|^2$ and $|1-S|^2$ by their energy averages.
The philosophy of the optical model is to address these energy-averaged
cross sections directly, without explicitly computing the fluctuations
of $S$ around its mean. This motivates identifying the \textit{optical}
(or direct) S-matrix with the smoothly varying average itself,
$S^{\rm opt}\equiv\langle S\rangle$, and constructing a potential whose elastic S-matrix
reproduces $S^{\rm opt}$.  It can be shown that 
this potential is the energy-averaged optical potential of Sect.~\ref{sec:optical} \citep[see][]{MahauxWeidenmuller1969}.  Equivalently, one simply splits
\begin{equation}
  \mathbf{S}(E) = \langle\mathbf{S}(E)\rangle
                + \bigl[\mathbf{S}(E)-\langle\mathbf{S}(E)\rangle\bigr]
  \label{eq:decomp}
\end{equation}
into a smooth part  (the
direct amplitude) and a rapidly fluctuating remainder.

Having
$\langle S\rangle$ is not the end of the story, since
$\langle|S|^2\rangle\neq|\langle S\rangle|^2$ in general. Instead,
\begin{equation}
  \langle|S|^2\rangle = |S^{\rm opt}|^2 + \bigl(\langle|S|^2\rangle-|\langle S\rangle|^2\bigr),
  \qquad
  \langle|1-S|^2\rangle = |1-S^{\rm opt}|^2 + \bigl(\langle|S|^2\rangle-|\langle S\rangle|^2\bigr),
\end{equation}
so that
\begin{equation}
  \langle\sigma_r\rangle = \sigma_r^{\rm opt}-\sigma^{\rm fl},
  \qquad
  \langle\sigma_{\rm el}\rangle = \sigma_{\rm el}^{\rm opt}+\sigma^{\rm fl},
  \label{eq:optfl}
\end{equation}
with the optical (direct) cross sections
$\sigma_r^{\rm opt}=\tfrac{\pi}{k^2}(1-|S^{\rm opt}|^2)$,
$\sigma_{\rm el}^{\rm opt}=\tfrac{\pi}{k^2}|1-S^{\rm opt}|^2$, and the
fluctuation, or compound, term
\begin{equation}
  \sigma^{\rm fl} = \frac{\pi}{k^2}\bigl(\langle|S|^2\rangle-|\langle S\rangle|^2\bigr).
  \label{eq:sigmafl}
\end{equation}
This is precisely the decomposition already anticipated in
Eq.~\eqref{eq:decomp}: $\sigma^{\rm fl}$, the compound elastic cross
section, is associated with formation of a compound nucleus that decays
back to the entrance channel, a process that lives for a time
$\tau_{\rm CN}=\hbar/\Gamma\sim10^{-18}$–$10^{-16}$~s — long enough to
lose all memory of the entrance channel~\citep{FriedmanWeisskopf1955} —
and it is experimentally
indistinguishable from genuine elastic scattering, so it must be
included in the measured $\langle\sigma_{\rm el}\rangle$. No interference
cross term survives the average~\citep{Feshbach1954}: this incoherence,
Eq.~\eqref{eq:decomp}, generalizes immediately to any pair of channels,
\begin{equation}
  \langle\sigma_{0a}\rangle = \sigma_{0a}^{\rm opt}+\sigma_{0a}^{\rm fl},
  \qquad
  \sigma_{0a}^{\rm opt} = \frac{\pi}{k^2}|\delta_{0a}-S_{0a}^{\rm opt}|^2,
  \qquad
  \sigma_{0a}^{\rm fl} = \frac{\pi}{k^2}\bigl(\langle|S_{0a}|^2\rangle-|\langle S_{0a}\rangle|^2\bigr),
  \label{eq:incoherent}
\end{equation}
with a direct-reaction model needed to compute $S_{0a}^{\rm opt}$, and,
as developed in Sect.~\ref{sec:compound} below, a statistical model
needed for $\sigma_{0a}^{\rm fl}$.

\subsection{The optical background: separating $H_F$ from $H_B$}
\label{sec:optbg}
In order to  connect the separation of the $S$ matrix into a smooth and a fluctuating part with the underlying structure, we will implement here the derivation of the optical background formalism of Kawai, Kerman, and McVoy~\citeyearpar{Kawai1973}.   
Let us first define the (Lorentzian)
energy average of the exact $P$-space wavefunction over an interval $I$,
\begin{equation}
  \bar\psi_P(E) \equiv \langle P\psi\rangle_I \simeq P\psi(E+iI/2),
  \label{eq:psibar}
\end{equation}
with $I$ chosen in the window $D\ll I\ll\Gamma_\downarrow$ (see
Sect.~\ref{S2.2.3} and Fig. \ref{fig:fragmentation}) — wide enough to average over many $H_F$
resonances, narrow enough not to wash out the collective width itself.
Substituting $E\to E+iI/2$ into Eq.~\eqref{eq:P-eff} shows that
$\bar\psi_P$ obeys its own effective Schr\"odinger equation,
\begin{equation}
  \bigl[E-H_{PP}-\Sigma(E+iI/2)\bigr]\bar\psi_P = 0,
  \label{eq:optaveraged}
\end{equation}
i.e., $\bar\psi_P$ is generated by exactly the optical potential
$U_0+U_{\rm pol}$ of Sect.~\ref{sec:optical}, only continued off the
real axis by the averaging width $I$. Define the corresponding effective
Hamiltonian and Green's function,
\begin{equation}
  \mathcal H_P^{\rm opt}(E) \equiv H_{PP}+\Sigma(E+iI/2),
  \qquad
  \mathcal G_P^{\rm opt}(E) \equiv \bigl[E-\mathcal H_P^{\rm opt}(E)\bigr]^{-1}.
  \label{eq:Hopt}
\end{equation}

We want to rewrite the exact equation for
$P\psi$, Eq.~\eqref{eq:P-eff}, with $\bar\psi_P$ itself as the driving
term. The resolvent identity
\begin{equation}
  G_Q(E)-G_Q(E+iI/2) = G_Q(E)\,\frac{iI}{2}\,G_Q(E+iI/2) 
  \label{eq:resolventid}
\end{equation}
lets us split $\Sigma(E)$ of Eq.~\eqref{eq:P-eff} into an
optical-background piece and a remainder,
\begin{equation}
  \Sigma(E) = \Sigma(E+iI/2) + V_{PQ}(E)\,G_Q(E)\,V_{QP}(E),
  \label{eq:split}
\end{equation}
where the modified couplings are defined as
\begin{equation}
  V_{PQ}(E) \equiv H_{PQ}\left[\frac{iI/2}{E-H_{QQ}+iI/2}\right]^{1/2},
  \qquad
  V_{QP}(E) \equiv \left[\frac{iI/2}{E-H_{QQ}+iI/2}\right]^{1/2}H_{QP},
  \label{eq:VPQ}
\end{equation}
 Substituting
Eq.~\eqref{eq:split} into Eq.~\eqref{eq:P-eff} we get,
\begin{equation}
  \bigl[E-\mathcal H_P^{\rm opt}(E)\bigr]P\psi = V_{PQ}(E)\,G_Q(E)\,V_{QP}(E)\,P\psi,
  \label{eq:twopotential}
\end{equation}
whose general solution, with $\bar\psi_P$ as the homogeneous piece, is
\begin{equation}
  P\psi = \bar\psi_P + \mathcal G_P^{\rm opt}(E)\,V_{PQ}(E)\,
  \bigl[E-H_{QQ}-V_{QP}(E)\mathcal G_P^{\rm opt}(E)V_{PQ}(E)\bigr]^{-1}
  V_{QP}(E)\,\bar\psi_P.
  \label{eq:PspaceOpt}
\end{equation}
This has exactly the structure of Eq.~\eqref{eq:PspaceLSexact}, with the
bare channel wavefunction $\xi_0$ replaced by the smooth $\bar\psi_P$,
and the bare couplings $H_{PQ},H_{QP}$ replaced by the damped
$V_{PQ},V_{QP}$. Expanding the bracketed operator in its biorthogonal
eigenstates $|q\rangle$, eigenvalues $\mathcal E_q$, and neglecting the
residual closed-channel continuum gives
\begin{equation}
  P\psi = \bar\psi_P + \mathcal G_P^{\rm opt}(E)\sum_q
  \frac{V_{PQ}(E)|q\rangle\langle q|V_{QP}(E)}{E-\mathcal E_q}\,\bar\psi_P.
  \label{eq:PspaceOptPoles}
\end{equation}

The S-matrix is obtained from the asymptotic form of
Eq.~\eqref{eq:PspaceOptPoles} 
\begin{equation}
  S_{cc'}(E) = \bar S_{cc'}(E) - i\sum_q\frac{g_{qc}(E)g_{qc'}(E)}{E-\mathcal E_q},
  \qquad
  g_{qc}(E) = \sqrt{2\pi}\sum_{c'}\langle\bar\psi_{c'}(E)|V_{c'Q}(E)|q\rangle,
  \label{eq:Sbackground}
\end{equation}
where $\bar S_{cc'}(E)$ is the S-matrix generated by
$\mathcal H_P^{\rm opt}(E)$ alone,  and the second term, representing the fluctuating (compound) part, has \emph{exactly zero-average} by construction. 

The \emph{direct} part of the S-matrix
$\bar S_{cc'}(E)$  is not just formally smooth — it has the
same structure as an ordinary DWBA matrix element, 
\begin{equation}
  \bar S_{cc'}(E) \;\propto\; \langle\bar\psi_{c}|\Delta V|\bar\psi_{c'}\rangle
  \label{eq:barSDWBA}
\end{equation}
where $\Delta V$ is an energy-independent coupling, specific to the actual pair of channels $c$ and $c'$ coupled by a direct reaction. The  wavefunctions $\bar\psi_{c}$  are generated by the very same energy-averaged complex optical potential
$\langle V(E)\rangle=V(E+iI/2)$ of Sect.~\ref{sec:optical}, generalized for the many-channel case. In other
words, the smooth background is not merely ``whatever is left once the
resonances are subtracted off'': it is positively identified with the direct-reaction amplitude according to the DWBA  (see Sect.~\ref{sec:dwba}).

% \textit{The central result.} Equation~\eqref{eq:Sbackground} looks the
% same as the naive sum-over-poles form we started from in
% Eq.~\eqref{eq:T-Feshbach}, but it is not: by construction, $\bar\psi_P$
% — and hence $\bar S_{cc'}$ — is already stationary under further
% averaging over $I$, $\langle\bar\psi_P\rangle_I\simeq\bar\psi_P$. Since
% the reaveraged $\langle S_{cc'}\rangle_I$ must reproduce this stable
% background exactly, the residual pole sum in Eq.~\eqref{eq:Sbackground}
% is \textit{forced} to average to zero on its own — not because its
% phases are assumed random, but because the square root in $V_{QP}(E)$
% sweeps the phase of each term through a full $2\pi$ as $E$ crosses the
% corresponding pole $\mathcal E_q$. This is precisely the identity
% anticipated, without proof, in Eq.~\eqref{eq:decomp} below: it identifies
% $\bar S_{cc'}(E)$ with the optical S-matrix $S^{\rm opt}$ of
% Sect.~\ref{sec:decomp}, ties the averaging interval used there to the
% same Lorentzian width $I$ introduced in Eq.~\eqref{eq:psibar}, and — this
% is the payoff — supplies the general, model-independent criterion that
% Figure~\ref{fig:iar} needed outside information (isospin selection
% rules) to supply for that one case: genuine $H_B$ structure is whatever
% survives in $\bar S_{cc'}$, built from a wavefunction provably stable
% under re-averaging, not merely whatever happens to look smooth in one
% coarse-resolution measurement.

We can now close the loop opened in Sect.~\ref{sec:properties}. The
smooth background $\bar S_{cc'}(E)$, generated by
$\mathcal H_P^{\rm opt}(E)=H_{PP}+\Sigma(E+iI/2)$ and carrying no memory
of individual $Q$-space eigenstates, is the S-matrix incarnation of collective $H_B$ physics (thick lines in Figure~\ref{fig:fragmentation}, see Sect. \ref{sec:direct}); the
residual sum over the densely spaced poles $\mathcal E_q$ that remain
once that coherent piece has been subtracted off is the contribution of
the  GOE-distributed eigenstates of $H_F$. The averaging interval $I$ is the
resolution knob of Figure~\ref{fig:iar}: chosen in the window
$D\ll I\ll\Gamma_\downarrow$ of Sect.~\ref{S2.2.3}, it separates
one from the other,  with the residual fluctuation  averaging to zero. In Sects.~\ref{S5.3}
and~\ref{sec:direct} we briefly connect what we have just found to the standard approaches for the modeling of compound and direct reactions, respectively. 

\subsection{Modeling Compound Reactions}\label{S5.3}
\label{sec:compound}
 
\subsubsection{The Statistical Model: Hauser-Feshbach}
\label{sec:HF}

While the optical cross sections follow from the single-particle
Schrödinger equation with the optical potential, $\sigma^{\rm fl}$
requires a model for the statistical properties of the S-matrix
fluctuations — one of the central problems of nuclear reaction theory,
tackled by Weisskopf, Feshbach, Bloch, Kerman, Moldauer, Weidenm\"uller
and others~\citep{WeisskopfEwing1940,Feshbach1954,Bloch,Brown1959,Moldauer1964,MahauxWeidenmuller1969,Kawai1973,MahauxWeidenmuller1979}. The most widely used model is Hauser-Feshbach
theory~\citeyearpar{HauserFeshbach1952}, built on Bohr's independence
hypothesis~\citeyearpar{Bohr1936} that the compound nucleus decays independently
of how it was formed. We can now derive this explicitly from the
statistics of the resonance sum, rather than simply postulate it,
closing the loop with the compound-nucleus chaos of
Sect.~\ref{sec:properties}.

By Sect.~\ref{sec:optbg}, the fluctuating remainder
$S_{0a}-\langle S_{0a}\rangle$ is exactly the resonance sum carried over
from Eq.~\eqref{eq:T-diagonal},
\begin{equation}
  S_{0a}(E)-\langle S_{0a}(E)\rangle
  = -2\pi i\sum_\mu \frac{g_{a\mu}(E)\,g_{0\mu}(E)}
  {E-E_\mu(E)+\tfrac{i}{2}\Gamma_\mu(E)}.
  \label{eq:Sfluct}
\end{equation}
Bohr's hypothesis, made statistically precise, is the assumption that in
a chaotic compound nucleus the coupling amplitudes $g_{a\mu}$ of
different resonances $\mu$ are uncorrelated random variables (so
$\mu\neq\mu'$ cross terms vanish on averaging), and that, for a genuine
rearrangement channel $a\neq0$, $g_{a\mu}$ and $g_{0\mu}$ themselves are
statistically independent for the same resonance:
\begin{equation}
  \overline{g_{a\mu}\,g_{0\mu}\,g_{a\mu'}^*g_{0\mu'}^*}
  = \delta_{\mu\mu'}\,\overline{|g_{a\mu}|^2}\;\overline{|g_{0\mu}|^2},
  \qquad a\neq0.
  \label{eq:randomphase}
\end{equation}
Squaring Eq.~\eqref{eq:Sfluct} and averaging under this assumption
collapses the double sum over $\mu,\mu'$ to a single diagonal sum,
\begin{equation}
  \overline{|S_{0a}-\langle S_{0a}\rangle|^2}
  = \sum_\mu \frac{\overline{\Gamma_0}\,\overline{\Gamma_a}}
  {(E-E_\mu)^2+\Gamma^2/4},
\end{equation}
where $\overline{\Gamma_c}\equiv2\pi\overline{|g_{c\mu}|^2}$ and $\Gamma$ is the
local average total width, since widths vary slowly compared to the
spacing between the many resonances being summed. Converting that sum
into an energy integral, 
($\sum_\mu\to\int dE_\mu/D$) and using the standard Lorentzian integral
$\int dE_\mu\,[(E-E_\mu)^2+\Gamma^2/4]^{-1}=2\pi/\Gamma$ gives
\begin{equation}
  \overline{|S_{0a}-\langle S_{0a}\rangle|^2}
  = \frac{2\pi\,\overline{\Gamma_0}\,\overline{\Gamma_a}}{D\,\Gamma}.
\end{equation}
Identifying the transmission coefficient with the average reduced width
in the usual way, $T_c\equiv2\pi\overline{\Gamma_c}/D$, and using
$\Gamma=\sum_c\overline{\Gamma_c}$, the level spacing $D$ cancels, and
Eq.~\eqref{eq:sigmafl} finally gives
\begin{equation}
  \sigma_{0a}^{\rm fl} = \frac{\pi}{k^2}\,\frac{T_0T_a}{\sum_c T_c},
  \label{eq:HF}
\end{equation}
Bohr's independence hypothesis, now a consequence of
chaotic, Porter-Thomas-distributed resonance couplings rather than an
assumption in its own right. 

This generalizes immediately, upon resolving channels by total angular
momentum $J$ and parity $\pi$, to $\sigma_{0a}(E)=\tfrac{\pi}{k_0^2}
\sum_{J\pi}(2J+1)\,T_0^{J\pi}T_a^{J\pi}/\sum_cT_c^{J\pi}$, with the
transmission coefficients defined in terms of the optical S-matrix,
\begin{equation}
  T_c = 1-|S_{cc}^{\rm opt}|^2,
  \label{eq:Tc}
\end{equation}
computed directly from the optical model of Sect.~\ref{sec:optical}
by solving the elastic scattering problem in channel $c$ alone.
Equation~\eqref{eq:HF} has the transparent form of a two-step process:
formation of the compound nucleus in channel $0$ (probability
$\propto T_0$), branching among all possible decays (probability
$T_a/\sum_cT_c$). It remains the workhorse for compound-nucleus cross
sections in nuclear reaction codes, from evaluated nuclear data to
large-scale reaction-network calculations in nuclear astrophysics, though
the limits of its underlying statistical assumptions remain a matter of
active research. An alternative approach models the S-matrix
fluctuations directly via random matrix theory, usually within the
Gaussian Orthogonal Ensemble already invoked in
Sect.~\ref{sec:properties}.

 Equation~\eqref{eq:HF} is a
closed, predictive scheme once three ingredients are supplied:
optical-model transmission coefficients $T_c$ for particle channels, $\gamma$-ray
transmission coefficients from a photon-strength function, and the
nuclear level density $\rho^{J\pi}(E)$  (refined
with spin-cutoff and back-shift corrections, Sect.~\ref{sec:intro})
for every residual nucleus reachable by particle, fission, or $\gamma$ emission.

 Equation~\eqref{eq:HF} rests
on an incoherent average of $\Gamma_{\mu 0}\Gamma_{\mu a}$ over
resonances $\mu$, which is only exact if the fluctuations of individual
partial widths around their mean are neglected. In reality partial
widths obey Porter-Thomas ($\chi^2$ with one degree of freedom)
statistics, and $\overline{\Gamma_0\Gamma_a}\neq
\overline{\Gamma_0}\,\overline{\Gamma_a}$ in general: the correlation is
strongest exactly in the elastic channel ($a=0$), where the width
fluctuation is not averaged away by summing over many final states as it
is for genuine rearrangement channels. This produces an \textit{elastic
enhancement factor} — the true elastic cross section exceeds the naive
Eq.~\eqref{eq:HF} by a factor of order $2$--$3$ in the limit of a single
open channel, decreasing toward $1$ as more channels open and the
central-limit theorem takes over. Moldauer's
correction~\citeyearpar{Moldauer1964} and the
Hofmann-Richert-Tepel-Weidenm\"uller (HRTW) scheme~\citeyearpar{Hofmann1975}
supply the appropriate correction factors, derived from random-matrix
theory applied directly to the level matrix $D(E)$ of
Eq.~\eqref{eq:Dmatrix}; modern Hauser-Feshbach codes include them as
standard.

\subsection{Modeling Direct Reactions}
\label{sec:direct}

% \subsubsection{The Optical Model}

% The energy-averaged optical potential derived in Sect.~\ref{sec:optical} plays a central role in the description of direct nuclear reactions. The potential entering everything that follows
% in this section, however, is not $\Sigma(E)$ evaluated on the real
% axis, but its energy-averaged form of Sect.~\ref{sec:optbg},
% $\mathcal H_P^{\rm opt}(E)=H_{PP}+\Sigma(E+iI/2)$ of Eq.~\eqref{eq:Hopt}
% — the object actually extracted from elastic-scattering data, itself
% necessarily an energy-averaged observable (Sect.~\ref{sec:decomp}),
% and the one that generates the distorted waves used throughout DWBA and
% coupled channels below.

\subsubsection{The Distorted-Wave Born Approximation (DWBA)}
\label{sec:dwba}

 Under
the natural statistical assumption of Sect.~\ref{sec:optbg} — that
the coupling amplitudes $g_{a\mu}$ of different channels are uncorrelated
for a shared compound resonance $\mu$ — the ensemble-averaged S-matrix
is diagonal: $\langle S_{ab}\rangle=0$ for $a\neq b$. Since we have assumed that $H_{PP}$ is diagonal in channel space (see the discussion at the end of Sect.~\ref{sec:Feshbach}), no direct
reaction would populate any channel but the entrance one. This is
flatly contradicted by experiment: forward-peaked angular distributions
with a smooth, mild energy dependence, well reproduced by the DWBA
machinery of the present section, are the rule rather than the exception
for a large class of inelastic and transfer channels. The independence
assumption must therefore fail for exactly those channels — and
understanding why is the first order of business before using DWBA at
all.

The channels for which it fails are precisely those tied to collective,
$H_B$-type excitations. Consider, for example, a neutron scattering off the target
nucleus of $A-1$ nucleons of Sect.~\ref{sec:Feshbach}, exciting either
the elastic channel, $|n(E)\rangle\otimes|(A-1)(0)\rangle$ (channel
$0$), or the first collective quadrupole surface vibration of the
target, $|n(E-E_2)\rangle\otimes|(A-1)(E_2)\rangle$ (channel $a$), with
$E_2=\hbar\omega_2^{\rm vib}$ of Eq.~\eqref{eq:vib}. Both the ground
state $|(A-1)(0)\rangle$ and the one-phonon state $|(A-1)(E_2)\rangle$
are built, in RPA, from a coherent superposition of 
particle-hole excitations around the Fermi surface — the ground state
is the phonon vacuum, the excited state a single phonon on top of it~\citep{RingSchuck1980,Rowe1970}.
Picture Figure~\ref{fig:fragmentation}b) drawn twice, once for each
channel's overlap with the compound spectrum of the $A$-body system
(the thick lines now representing $|g_{0\mu}|$ and $|g_{a\mu}|$,
respectively): because the two entrance configurations are built from
the same particle-hole raw material, they fragment into the compound
background with very similar centroids and spreading widths
$\Gamma_\downarrow$.  This is the
microscopic content of the ``chosen channels'' of Sano, Yoshida, and
Terasawa~\citeyearpar{Sano1958}: $g_{0\mu}$ and $g_{a\mu}$ are correlated, $\overline{g_{0\mu}g_{a\mu}^*}\neq0$, and the
corresponding off-diagonal element $\langle S_{0a}\rangle$ survives the
average. When this happens, we will say that the channels $0$ and $a$ are \emph{directly coupled}.

Let us now provide a more explicit justification for Eq.~\eqref{eq:barSDWBA}. The effective, averaged-out Hamiltonian $\mathcal H_P^{\rm opt}(E)$ of Eq.~\eqref{eq:Hopt} is an operator defined in channel space with non-zero matrix elements $\mathcal {H}_{cc'}^{\rm opt}(E)$ between all channels $c$ and $c'$ that happen to be correlated in the way described above. We define the distorted waves $\bar\psi^{(0)}_0$ and $\bar\psi^{(0)}_a$,
\begin{equation}
  \bigl[E-\mathcal H_{00}^{\rm opt}(E)\bigr]\bar\psi^{(0)}_0 = 0, \qquad
  \bigl[E-\mathcal H_{aa}^{\rm opt}(E)\bigr]\bar\psi^{(0)}_a = 0.
  \label{eq:distortedwaves}
\end{equation} 
In this restricted two-channel space, the wavefunction $\bar\psi_P=\bar\psi_0 + \bar\psi_a$ of Eq.~\eqref{eq:PspaceOpt} satisfies
\begin{equation}
  \bigl[E-\mathcal H_P^{\rm opt}(E)\bigr]\bar\psi_P = 0, \qquad
  \mathcal H_P^{\rm opt}(E) = \begin{pmatrix}
    \mathcal H_{00}^{\rm opt}(E) & \mathcal H_{0a}^{\rm opt}(E) \\
    \mathcal H_{a0}^{\rm opt}(E) & \mathcal H_{aa}^{\rm opt}(E)
  \end{pmatrix}.
  \label{eq:optmatrix}
\end{equation}
So,
\begin{align}\label{eq:2couple}
  \bigl[E-\mathcal H_{00}^{\rm opt}(E)\bigr]\bar\psi_0
  &= \mathcal H_{0a}^{\rm opt}(E)\,\bar\psi_a, \\
  \nonumber \bigl[E-\mathcal H_{aa}^{\rm opt}(E)\bigr]\bar\psi_a
  &= \mathcal H_{a0}^{\rm opt}(E)\,\bar\psi_0.
\end{align}
In order to address the measurement of channel $a$, we solve for the associated wavefunction,
\begin{align}
  \bar\psi_a
  &=\bigl[E^+-\mathcal H_{aa}^{\rm opt}(E)\bigr]^{-1} \mathcal H_{a0}^{\rm opt}(E)\,\bar\psi_0=  \mathcal G_{aa}^{\rm opt}(E)\,\mathcal H_{a0}^{\rm opt}(E)\,\bar\psi_0.
\end{align}
The DWBA now consists in approximating 
\begin{align}\label{eq:aproxdwba}
  \bar\psi_0\approx \bar\psi^{(0)}_0,
\end{align}
with which one can immediately obtain the T-matrix $T_{a0}$ from the asymptotic behaviour of (\ref{eq:aproxdwba})
\begin{align}\label{eq:Tdwba}
  T_{a0}= \langle\bar\psi^{(0)}_a|\mathcal H_{a0}^{\rm opt}|\bar\psi^{(0)}_0\rangle,
\end{align}
where $\mathcal H_{a0}^{\rm opt}$ can be assumed to be essentially energy-independent for an energy range of the order of $\Gamma_{\downarrow}$.
 It is worth stressing again what the two
distorted waves $\bar\psi^{(0)}_0,\bar\psi^{(0)}_a$ actually are: not free waves,
not waves in some auxiliary or channel-specific potential, but
solutions generated by the  energy-averaged, complex optical potential
of Sect.~\ref{sec:optical}. Therefore, Eq.~\eqref{eq:Tdwba} is a genuine DWBA T-matrix, where all the energy-dependence is contained in the initial and final distorted waves. 

The central object of direct reaction theory is, arguably, the potential $\mathcal H_P^{\rm opt}$. Its diagonal parts provide the distorted waves (\ref{eq:distortedwaves}), and the off diagonal elements provide the transition potentials between the directly coupled channels. However $\mathcal H_P^{\rm opt}$ is rarely  computed explicitly according to Sect. \ref{sec:optical}. Instead, the diagonal part is usually fitted from elastic scattering experiments, and  the off-diagonal,  transition-inducing interaction is modeled according to the specific process under study - a deformed mean-field term for
inelastic excitation of rotational bands,  the single-particle mean field in a transfer reaction, or a particle-vibration coupling in the collective model in the case of inelastic excitation of surface vibrations such as the example presented here~\citep{Jackson1970,Satchler1990,ThompsonNunes2009}.

DWBA is a first-order theory: it is reliable when $\mathcal H_{0i}^{\rm opt}$ is genuinely a
weak perturbation on the elastic channel, i.e., when the coupling to the
inelastic or transfer channel returns only a small fraction of the incident
flux. This holds well for weakly collective states and for transfer
reactions with small cross sections relative to the reaction total, but it
must be abandoned when the coupling itself is strong enough to alter the
elastic channel appreciably — precisely the case for the low-lying
collective vibrations and rotations singled out in Sect.~\ref{sec:intro}
as the hallmark of the collective picture.

\subsubsection{Coupled-Channels Methods}
\label{sec:cc}

When a small number of channels are coupled strongly — a rotational band
built on a well-deformed ground state, or a low-lying vibrational phonon
with large deformation parameter $\beta_\lambda$ — first-order perturbation
theory is no longer adequate: flux oscillates back and forth between the
elastic channel and the strongly coupled excited states many times before
the projectile leaves the interaction region, and DWBA, which allows flux
to leave the elastic channel only once, systematically underpredicts the
elastic depletion and misrepresents the excitation of the strongly coupled
state itself.

The coupled-channels method~\citep{Tamura1965} solves this problem non-
perturbatively by solving the set  of coupled radial
equations for the wavefunctions $u_\alpha(r)$ in every channel $\alpha$
retained explicitly (ground state, band members, phonon multiplets\dots). This is obtained by projecting sets such as (\ref{eq:2couple}) over the states $\phi_\alpha(\xi)$ of the target nucleus associated with each one of the selected channels, 
\begin{equation}
  \left[\frac{d^2}{dr^2} + k_\alpha^2
  - \frac{l_\alpha(l_\alpha+1)}{r^2}
  - \frac{2\mu}{\hbar^2}\,U_{\alpha\alpha}(r)\right] u_\alpha(r)
  = \frac{2\mu}{\hbar^2}\sum_{\beta\neq\alpha}
  U_{\alpha\beta}(r)\,u_\beta(r),
  \label{eq:CC}
\end{equation}
with diagonal potentials $U_{\alpha\alpha}$  and off-diagonal couplings
$U_{\alpha\beta}$ generated from the corresponding matrix elements of $\mathcal H_{a0}^{\rm opt}$ between the states  $\phi_\alpha(\xi),\phi_\beta(\xi)$.  DWBA is recovered exactly as the first Born
iteration of Eq.~\eqref{eq:CC} in the weak-coupling limit. Because
Eq.~\eqref{eq:CC} is solved to all orders, reorientation effects
(the excited state coupling back to the elastic channel and to other
excited states, not just being fed by it once) and multi-step transfer
paths are included automatically, at the cost of a much larger numerical
problem.

% A striking application is sub-barrier heavy-ion fusion, where couplings to
% low-lying collective modes of projectile and target split the single
% barrier of the bare (uncoupled) potential into a distribution of barriers.
% This distribution is recovered experimentally as the second energy
% derivative of $E\sigma_{\rm fus}(E)$~\cite{Rowley1991}, and reproducing its
% detailed shape and multi-peaked structure has become one of the most
% sensitive tests of coupled-channels reaction theory against nuclear
% structure input, directly linking the fusion cross section — an
% integrated, seemingly structureless observable — to the collective
% spectrum of Sect.~\ref{sec:intro}. When transfer channels are included on
% the same non-perturbative footing as inelastic excitations, the method is
% usually called coupled-reaction-channels (CRC), and is the framework
% underlying essentially all modern quantitative analyses of nucleon-transfer
% angular distributions.

\subsubsection{Connection to the Collective Picture}
\label{sec:collective}

The purpose of the machinery assembled in Sects.~\ref{sec:optical}--
\ref{sec:cc} is not merely to reproduce cross sections: it is to use direct
reactions as a spectroscopic tool, turning a measured angular distribution
into a quantitative statement about the elementary modes of excitation
introduced in Sect.~\ref{S2.2.3}. 

Inelastic scattering and Coulomb excitation, analyzed with the coupling
potentials of Eq.~\eqref{eq:CC}, measure the reduced transition
probabilities $B(E\lambda)$ that quantify the collectivity of a vibrational
or rotational state — the direct experimental counterpart of the
vibrational and rotational energy scales estimated in
Eqs.~\eqref{eq:vib}--\eqref{eq:rot}. One-nucleon transfer reactions,
analyzed in DWBA or CRC via Eq.~\eqref{eq:Tdwba}, measure spectroscopic
factors that quantify how much of a given
single-particle orbital's strength survives as a single, identifiable
state rather than being fragmented into the compound background of
Sect.~\ref{sec:compound} — a direct experimental measurement of the
elementary-mode fragmentation discussed in Sect.~\ref{S2.2.3}.

Two-nucleon transfer reactions, such as $(p,t)$ and $(t,p)$, occupy a
special place in this picture: because the transferred pair is
predominantly coupled to $J^\pi=0^+$, their cross sections are enhanced
over the single-particle estimate by a factor directly proportional to the 
pairing correlations of the initial and final states, providing the most
direct reaction-based probe of nuclear superfluidity introduced in
Sect.~\ref{S2.2.2}~\citep{Broglia2021,Potel2013}. Ground-state-to-ground-state $(p,t)$ cross
sections in superfluid nuclei are enhanced by one to two orders of
magnitude over an uncorrelated estimate, while transitions to the
pairing-vibrational states built on top of the ground state carry the
complementary strength — the reaction-theory manifestation of the
pairing-vibration and pairing-rotation spectra.
 
% In every case the underlying logic is the same: the reaction amplitude
% factorizes (Eq.~\eqref{eq:DWBA}) into a reaction part, computed once and
% for all from the optical potentials of Sect.~\ref{sec:optical}, and a
% structure part — a form factor, a transition density, a pair-transfer
% amplitude — that is exactly the elementary-mode content of $\langle
% \mathbf{S}(E)\rangle$ identified in Sect.~\ref{S2.2.3}. 

Direct reactions are the experimental manifestation that makes
the collective picture of Sect.~\ref{S2.2.2} apparent, and that testify to the soundness of the  framework presented in Sect.~\ref{S2.2.3}. Within this context, direct reactions give a quantitative answer to the question \emph{to what extent is a nucleus an object?}. In other words, they provide considerable insight into the onset of emergent properties in finite, strongly-interacting many-body quantum systems, in a regime which is far away from the $N\to\infty$ thermodynamic limit, and where phase transitions and associated order parameters can only be incipient \citep{Anderson2018}.

% ============================================================
\section{Outlook}
\label{sec:outlook}

The picture assembled in this chapter — two extreme, complementary limits
of the nuclear many-body spectrum, connected  through the analytic
structure of the S-matrix and modeled, in each limit, by the machinery of
Sects.~\ref{sec:compound} and~\ref{sec:direct} — is a mature framework,
refined over eight decades since Bohr's original compound-nucleus
hypothesis \citep{Bohr1936} and Breit and Wigner's resonance formula \citep{BreitWigner}. It is also the foundation on which the specialized chapters
of this volume build: the reader is referred to the companion chapters on
the optical model and global potential parametrizations, on
Hauser-Feshbach and pre-equilibrium codes for applied and evaluated
nuclear data, on $R$-matrix theory for resonance-region and low-energy
astrophysical reactions, and on coupled-channels and coupled-reaction-
channels codes for detailed transfer and inelastic-scattering analyses, each
of which develops one branch of the tree rooted in
the considerations made in the present Chapter.

Much has been omitted in this brief introduction. \textit{Ab initio} reaction
theory — extending no-core shell-model, coupled-cluster, and lattice
effective-field-theory methods, formulated for bound states, to include the
scattering continuum explicitly — aims to derive both the optical potential
and the compound-nucleus level density from a single microscopic
Hamiltonian, rather than fitting them independently as done throughout
Sects.~\ref{sec:compound}--\ref{sec:direct}~\citep{Navratil2016}.
 % \textit{Unified reaction
% theories} seek to describe, within one calculation, processes that
% interpolate between the pure direct and pure compound limits — a compound
% nucleus formed after a direct step, or a direct reaction proceeding through
% a doorway state with intermediate level density — of which the surrogate-
% reaction method for constraining otherwise inaccessible compound-nucleus
% cross sections~\cite{Escher2012} is a mature, actively used example.
\textit{Reactions near the drip lines}, where the projectile or target is a
weakly bound or halo nucleus, challenge the very separation between
structure and reaction assumed throughout this chapter: the breakup
channel is simultaneously a decay of the projectile's own structure and a
reaction channel that must be treated on the same non-perturbative footing
as inelastic and transfer channels in Eq.~\eqref{eq:CC}~\citep{Moro2025}. Finally,
\textit{nuclear astrophysics} — the calculation of radiative-capture and
transfer cross sections at energies far below any accessible beam energy,
central to modeling the $r$-process and other stellar nucleosynthesis
pathways — depends critically on every tool assembled here, extrapolated
into regimes where direct measurement is impossible and indirect methods,
validated against the exact S-matrix formalism of Sect.~\ref{sec:Smatrix},
are the only recourse.

The very premise of this Chapter — the use of a handful of basic facts about the nuclear many-body spectrum of typical nuclei to organize the entire apparatus of nuclear reaction theory in terms of two coexisting but seemingly contradictory pictures — also speaks for its limitations: exotic nuclei, with extreme neutron-to-proton ratios, low particle-separation energies, and very low level densities, may fail, in one way or another, to accommodate the general picture presented here. In particular, it will probably be necessary to explore the gray area between the two extremes of the compound and direct limits in order to understand some important reaction mechanisms in nuclei far from stability, such as the ones relevant for the description of the $r$-process.

We have tried to present the material in a way that emphasizes the role of energy resolution in determining the degrees of freedom that are relevant for the description of nuclear spectra and processes,
\begin{itemize}
  \item \textbf{high resolution} $(I\ll D)$: individual resonances,  $H_F$ Hamiltonian, \emph{compound reactions}; 
  \item \textbf{low resolution} $(D\ll I \ll \Gamma_{\downarrow})$: collective states, $H_B$ Hamiltonian, \emph{direct reactions},
\end{itemize}
where $D$ is the average level spacing between compound nucleus states, $I$ the energy resolution, and $\Gamma_{\downarrow}$ the spreading width of the collective states. This state of affairs is quite natural in the general context of quantum many-body systems, where the relevant degrees of freedom are dictated by the energy resolution with which the system is probed, and emergent properties arise when crossing different resolution realms. What is remarkable in the nuclear case is the degree of coexistence of such different regimes, and our ability to turn smoothly (both theoretically and experimentally) the two relevant knobs: the number of nucleons, and the energy resolution. The fact that these knobs are largely under experimental and theoretical control sets us in a privileged situation for the study of emergent phenomena in many-body systems.

The two nuclear pictures with which this chapter opened — the chaotic
Fermi gas and the ordered collective liquid drop — are not, in the end,
two different kinds of nuclei. They are two faces of the same many-body
system, and the entire apparatus of nuclear reaction theory is the set of
tools we have built to ask, of any given nucleus at any given energy,
which face it is showing us.

% ============================================================
\bibliographystyle{agsm}
\bibliography{refs}

\end{document}